\documentclass[
 reprint,
 amsmath,amssymb,
 aps,
]{revtex4-2}

\usepackage{graphicx}
\usepackage{bm}

\usepackage[colorlinks=true, linkcolor=blue,anchorcolor=blue, citecolor=blue,urlcolor=blue]{hyperref}

\usepackage[normalem]{ulem}
\usepackage{CJKutf8}
\begin{document}
\bibliographystyle{apsrev4-1}
\preprint{APS/123-QED}

\title{Chiral magnetism in lithium-decorated monolayer CrTe$_{2}$: Interplay between Dzyaloshinskii-Moriya interaction and higher-order interactions}
\author{Weiyi Pan$^{1}$}

\author{Changsong Xu$^{4,5}$}
\email{csxu@fudan.edu.cn}

\author{Xueyang Li$^{4,5}$}

\author{Zhiming Xu$^{1}$}

\author{Boyu Liu$^{4,5}$}

\author{Bing-Lin Gu$^{1,2}$}

\author{Wenhui Duan$^{1,2,3}$}

\affiliation{$^{1}$State Key Laboratory of Low Dimensional Quantum Physics and Department of Physics, Tsinghua University, Beijing 100084, China\\
$^{2}$Institute for Advanced Study, Tsinghua University, Beijing 100084, China \\
$^{3}$Frontier Science Center for Quantum Information, Beijing 100084, China\\
$^{4}$Key Laboratory of Computational Physical Science (Ministry of Education), State Key Laboratory of Computational Physical Science, and Department of Physics, Fudan University, Shanghai 200433, China\\
$^{5}$Shanghai Qi Zhi Institute, Shanghai 200030, China\\
}

\begin{abstract}
Chiral magnetic states in two-dimensional (2D) layered noncentrosymmetric magnets, which are promising advanced spintronic materials, are usually attributed to Dzyaloshinskii–Moriya interactions (DMI). However, the role of underlying higher-order spin couplings in determining the properties of chiral spin textures has much less reported. In this work, taking the lithium-decorated monolayer CrTe$_{2}$ (monolayer LiCrTe$_{2}$) as an example, we develop a first-principles-based comprehensive spin model constructed by using the symmetry-adapted cluster expansion method. Based on this spin model, we identify the ground state of monolayer LiCrTe$_{2}$ as a chiral spin spiral state, which can further assemble macroscopic chiral labyrinth domains (LD) under zero-field conditions as well as evolve into skyrmions under a finite magnetic field. Moreover, higher-order biquadratic and three-site interactions are identified to be responsible for modulating both the size and the field stability of the spin spiral state. Our study sheds light on complex magnetic couplings in 2D magnets.

\end{abstract}
\maketitle
\section{Introduction}
Novel chiral spin textures, such as chiral spin spirals (SS)\cite{intro1, intro2,intro3} and magnetic skyrmions\cite{skyrmion1,skyrmion2,skyrmion3}, have attracted much attention because they are fertile ground for both emergent magnetoelectric phenomenon explorations\cite{SkxME1,SkxME2,SkxME3,SkxME4} and advanced spintronic device applications\cite{SkXApp4,SkXApp2,SkXApp1}. The formation and basic characteristics (i.e., type, morphology and stability)\cite{STB1,STB2,STB3,STB4,STB5,DMIType,DMIType2,DMIType3,DMIType4} of these chiral spin textures are usually correlated with the asymmetric Dzyaloshinskii-Moriya interaction (DMI)\cite{DM3,DMI1,DMI2}, which favors a noncollinear spin arrangement. 
According to Moriya’s rule\cite{DMI1}, the key to inducing DMI is breaking the centrosymmetry of a system, which can be easily realized in two-dimensional (2D) magnetic materials\cite{2D1,2D2,2D3,2D4}. 
To date, many strategies for eliminating centrosymmetry in 2D magnetic systems have been theoretically and experimentally considered\cite{2D4,Janus1,Janus2,Janus3,E1,E2,E3,Moire1,Moire2,Moire3,Hetero1,Hetero2,Hetero3}. 

In addition to DMI, other spin interactions,   such as exchange frustration\cite{fru1,fru2,fru3}, magnetic anisotropy\cite{MAE1,Janus1}, and Kitaev interactions\cite{Janus1,Kitaev2}, can also have nontrivial effects on the basic properties of chiral spin configurations. Nevertheless, these spin couplings are all limited in the range of second-order interactions. Recently, higher-order spin interactions\cite{Novel1,Novel2,Novel3,Novel4,Novel5,Novel6,Novel7,Novel8,Novel9,Novel10, Novel11,Novel12,Novelexp1,Novelexp2,Novelexp3,Novelexp4}, 
have been predicted to exist in some realistic magnetic systems. Interestingly, higher-order interactions are proposed to play unique roles in the morphology and stability of chiral spin textures\cite{HOIeff,Novel3,HOIeff2,Novel4,HOIeff3}. One may be curious about whether any higher-order interactions exist in 2D magnets that could significantly influence the emergence of chiral spin textures. If such higher-order interactions exist, what do they look like? How do these higher-order interactions affect the fundamental characteristics of chiral spin structures? To answer these questions, a model system is required. 

A suitable system for addressing our concerns above is Li-decorated monolayer CrTe$_{2}$ (monolayer LiCrTe$_{2}$). When exfoliated from its bulk form\cite{ACT4,ACT6}, monolayer LiCrTe$_{2}$ is predicted to be energetically and dynamically stable\cite{ACT3}, with a significantly high FM transition temperature of approximately 200 K. Importantly, as Li ions are located on one side of CrTe$_{2}$, the centrosymmetry of the host system is disrupted, which potentially gives rise to a strong DMI. Although previous work has theoretically predicted DMI-induced chiral magnetic structures in monolayer LiCrTe$_{2}$\cite{ACT5}, the effect of underlying higher-order spin interactions on the fundamental characteristics of chiral spin textures remains unclear.

In this work, we reveal which and how higher-order interactions affect the chiral magnetic states of monolayer LiCrTe$_{2}$. Based on a newly developed symmetry-adapted cluster expansion (SACE) method, we construct a first-principles-based effective spin Hamiltonian. This Hamiltonian not only contains the commonly considered single-ion anisotropy (SIA) and two-body couplings, such as Heisenberg interaction and DMI,  but also accommodates symmetry-allowed higher-order spin couplings, which are beyond the framework of widely used second-order spin models. The resulting effective Hamiltonian predicts N$\Acute{\textup{e}}$el-like ferromagnetic SS as the magnetic ground state of monolayer LiCrTe$_{2}$. Among the higher-order spin interactions, biquadratic interactions and three-site interactions are identified to have nontrivial effects on the size and stability of SS. Further Monte Carlo (MC) simulations demonstrate the emergence of metastable N$\Acute{\textup{e}}$el-type skyrmions when considering an external magnetic field.

\section{METHODS}
\subsection{Computational workflow}
Our study proceeds through a systematic four-step process: Initially, we utilized the SACE method to generate all symmetry-allowed spin invariants based on the crystal structure of monolayer LiCrTe$_{2}$. Subsequently, employing density functional theory (DFT), we computed the energies of various randomly generated spin configurations. In the next stage, we employed a machine learning approach to discern the dominant terms from the myriad of possible invariants, thereby effectively reducing their number by fitting the parameters of the spin Hamiltonian. Finally, employing the derived Hamiltonian, we conducted Monte Carlo simulations, followed by a comprehensive analysis. Detailed descriptions of each step are provided below.

\subsection{Spin Hamiltonian 
from symmetry-adapted cluster expansion method}
The general form of the spin Hamiltonian is written as:
\begin{equation}\label{1}
    H = H_{0} + \sum_{\textup{N}} H_{\textup{N}}^{\textup{spin}}
\end{equation}
where the $\textit{H}_{0}$ refers to the nonmagnetic part, and $\textit{H}_{\textup{N}}^{\textup{spin}}$ indicates the N-order spin interaction terms. For example, the dominate second-order and fourth-order terms can be expressed as: 
 
\begin{equation}\label{2}
\begin{aligned}
       & H_{\textup{2-order}}^{\textup{spin}} = \sum_{\left \langle i,j \right \rangle} \sum_{\alpha, \beta} a_{ij}^{\alpha \beta} S_{i \alpha}S_{j \beta} \\
 &   H_{\textup{4-order}}^{\textup{spin}} = \sum_{\left \langle i,j,k,l \right \rangle} \sum_{\alpha, \beta, \gamma, \delta} a_{ijkl}^{\alpha \beta \gamma \delta} S_{i \alpha}S_{j \beta}S_{k \gamma} S_{l \delta}
\end{aligned}
\end{equation}.
 
Here the summation encompasses all potential spin components. The index $\left \langle ... \right \rangle $ refers to an M-body atomic cluster comprising magnetic atoms denoted as \textit{i}, \textit{j}, \textit{k}, and \textit{l}. The indices $\alpha$, $\beta$, $\gamma$, and $\delta$ denote spin components. For simplicity, the spins are normalized to $S$ = 1 in this study. Considering time-reversal symmetry, solely even-order terms are incorporated into this cluster-expansion Hamiltonian. Additionally, to balance accuracy and simplicity, higher-order terms in our model are truncated at the fourth order. Moreover, by regulating the cutoff distance between ions, our model considers only eight dominant clusters, encompassing a single-ion cluster, two-body clusters up to the third nearest neighbors, two three-body clusters, and two four-body clusters (refer to Fig. S4 in the Supplementary Material \cite{SM}). 

By further applying crystal symmetries, the equation (2) could be rewritten as:
\begin{equation}
\begin{aligned}
    &   H_{\textup{2-order}}^{\textup{spin}} = \sum_{n}L_{n} \sum_{m} a_{nm} S_{i_{m}\alpha_{m}}S_{j_{m}\beta_{m}} \\
    &   H_{\textup{4-order}}^{\textup{spin}} = \sum_{n}K_{n} \sum_{m} a_{nm} S_{i_{m}\alpha_{m}}S_{j_{m}\beta_{m}}S_{k_{m}\gamma_{m}}S_{l_{m}\delta_{m}}.
\end{aligned}    
\end{equation} 
where \textit{n} denotes each isolated invariant, with the summation conducted over \textit{m} to aggregate all terms within each invariant. It is evident that leveraging crystal symmetries significantly diminishes the number of terms, retaining only those permitted by symmetry. Consequently, this SACE Hamiltonian can theoretically encompass all permissible many-body and higher-order interactions, a scope beyond the confines of a second-order spin Hamiltonian framework.

This section provides technical insights into spin invariant generation using the SACE method. To enhance computational efficiency and model accuracy, it is imperative to truncate three key parameters: the number of interaction orders denoted as $N$, the number of sites $M$ encompassing spin interactions, and the maximum interaction distance $d_{M}$. Here, $d_{M}$ is defined such that the interaction distance between any two sites among the $M$ atoms is shorter than $d_{M}$. In this study, meticulous testing led to the adoption of $N=4$, $M=4$, $d_{2} = 9.0$ \AA, $d_{3} = 6.9$ \AA, and $d_{4} = 7.9$ \AA, ensuring comprehensive representation of dominant interaction patterns within the model while maintaining manageable computational costs.

\subsection{DFT calculations}
Having obtained the form of spin model, we hereby calculate the energies of various spin configurations with DFT. The energies of spin configurations would be used for fitting the spin model. We performed DFT calculations implemented in the Vienna $ab$ $initio$ simulation (VASP) with the projector augmented wave (PAW) method.  In our computational treatment, the generalized gradient approximation with the Perdew–Burke–Ernzerhof (PBE) functional was used\cite{PBE}. During the structural optimization, we fully relaxed both the atomic positions as well as the lattice constants of the system until the residual Hellmann-Feynman force per atom was less than 0.001 eV/\AA. Meanwhile, the electronic convergence criteria was chosen to be 10$^{-6}$ eV.  To treat the on-site Coulomb correlation effect of the Cr 3d orbitals, we set $U$ and $J$ values to be 3.0 eV and 0.6 eV, respectively, as used in the previous study\cite{JW}. A vacuum region of 20 \AA  \quad along the out-of-plane direction eliminate interaction between the periodic images. The vdW dispersion correction included in the DFT-D3 method was considered for accurately describing the structural properties\cite{DFT-D3}. 
For the concern of a unit cell, the cut-off energy of the plane wave was set to be 600 eV, with a $15\times15\times1$ Gamma-centered k-point mesh sampled in the Brillouin zone. After structural relaxation, the final lattice constant of monolayer LiCrTe$_{2 }$ we obtained is $a = b = 4.06 $\AA. 
To extract the energies of random spin configurations, a $5\times5\times1$ supercell was adopted. In this case, we carefully tested and reset the plane wave cut-off energy to be 400 eV, and a $\Gamma$-centered k-point mesh of $4\times4\times1$ was adopted, so that the accuracy can be maintained with less computational cost during the supercell calculations. During the calculations, the directions of the spins are constrained, while their magnitudes are fully relaxed. In order to fix the spin directions, we set LAMBDA = 10 in VASP, so that the spin would naturally locate near the initial configurations by minimizing the penalty energies. After DFT calculations, we deduct the penalty energy (\textless 0.05 meV/Cr) from the corresponding total DFT energies before fitting the model.

\subsection{Fitting the model with a machine learning approach}
The spin configurations as well as their DFT-calculated energies are used to fit the spin model, in order that we can obtain the values of $L_{n}$ and $K_{n}$, as well as $E_{0}$ of $H_{0}$. To better describe the fitting, we rewrite equation (1)-(3) as:
\begin{equation}
    E_{n}=E_{0}+\sum_{m}Y_{m}C_{n,m}
\end{equation}
where $E_{n}$ is the DFT total energy of the $n$-th spin configuration and $C_{n,m}$ are coefficients corresponded to the $m$-th invariants. 
The fitting was performed  with a machine learning method for constructing a Hamiltonian (MLMCH) as implemented in PASP\cite{PASP}. Details of this fitting method can be seen in Ref. \cite{LXY,XCS}. Such MLMCH method have already been used to study various magnetic systems\cite{Novel3, NovelNiI2, Novel2}.

In this study, the initial model is developed up to the four-body and fourth-order spin interactions, with distant neighbors extending up to the third nearest neighbors (3NN), while fully considering the spin-orbit coupling (SOC) effect. Subsequently, we conducted energy calculations for 575 random spin structures via DFT. Notably, within these 575 datasets, the training set comprises 444 sets of data, while the testing set comprises 131 sets. Leveraging the energies of these calculated spin structures, we repeated fit and refine the spin model within the MLMCH framework, ultimately achieving a model with minimal spin interaction terms and optimal numerical performance. For further details, refer to the Supplementary Material \cite{SM}.

\subsection{Monte Carlo simulations}
We carried out Parallel Tempering Monte Carlo (PTMC) simulations\cite{PTMC2} using the fitted Hamiltonian (1)-(3). Different sizes and shapes of supercells are adopted in our study. To be specific, the $40\times20\times1$ supercell (containing 1600 Cr ions) is used in Fig. \ref{2}, while a smaller  $30\times15\times1$ supercell (containing 900 Cr ions) is used in Fig. \ref{3} and Fig. \ref{4}. Both supercells are based on the rectangular cell defined by $\textbf{a}' = \textbf{a}$, $\textbf{b} = \textbf{b}' + 2\textbf{b}$, $\textbf{c}' = \textbf{c}$, where \textbf{a}, \textbf{b}, \textbf{c} are lattice vectors of the original parallelogram unit cell. During PTMC simulations, the initial magnetic configuration is randomly generated, and 120000 MC steps are performed for each temperature. To be specific, we set 400 replica exchange steps and 300 MC steps between every two replica exchange steps. The temperature is gradually cooled down from 315 K to the investigated lowest temperature, i.e., 2K.    
To further optimize the spin configuration after the MC simulations, a CG method \cite{CG}is applied so that the direction of each spin would rotate before the forces on each spin minimizes. In this case, the obtained magnetic state is located at its energy minima. The energy convergence criteria of this CG algorithm is set to be $10^{-6}$ eV. 

\section{RESULTS AND DISCUSSION}
\subsection{The effective model of monolayer LiCrTe$_{2}$}

\begin{figure*}[ht]
\includegraphics[scale = 0.36 ]{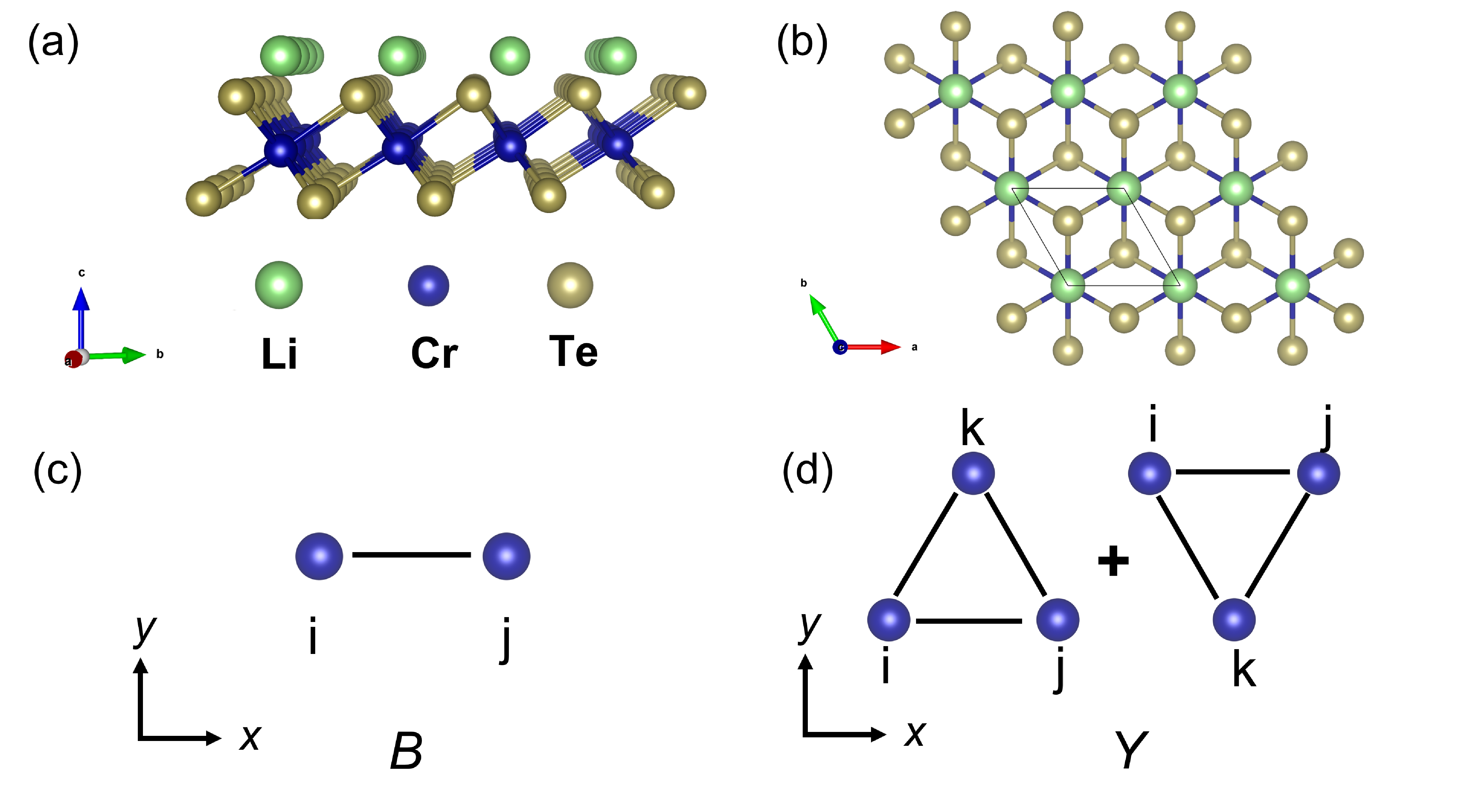}
\caption{\label{1} Crystal structure of monolayer LiCrTe$_{2}$. (a) Side view of monolayer LiCrTe$_{2}$. (b) Top view of monolayer LiCrTe$_{2}$. The parallelogram shows the in-plane unit cell. The atomic clusters involved in the biquadratic interaction and three-site interactions are shown in (c) and (d), respectively.
}
\end{figure*}
The crystal structure of monolayer LiCrTe$_{2}$ is shown in Figs. \ref{1}(a) and \ref{1}(b). The Cr atoms form a triangular lattice, and the nearest Cr-Cr distance is evaluated to be 4.06 \AA. It can be clearly seen that each Cr atom is surrounded by a Te octahedron, thus the resulting octahedron crystal field splits the 3d orbital of the Cr atoms into threefold degenerate t$_{2 g}$ levels and twofold degenerate e$_{g}$ levels. By locating a layer of Li atoms on the top of the monolayer CrTe$_{2}$, the inversion symmetry of the system is broken and the overall point group symmetry belongs to C$_{3v}$. Furthermore, the valence state of Cr changes from Cr$^{4+}$ in pristine monolayer CrTe$_{2}$ to 
 Cr$^{3+}$ in Li-decorated monolayer CrTe$_{2}$. In the latter case, the spin-up channel t$_{2 g}$ is occupied by 3 electrons, while the spin-down channel is empty. Furthermore, no electrons occupy the e$_{g}$ level. This electronic distribution results in a total spin magnetic moment of S = 3 $\mu_{B}$ for each Cr$^{3+}$ ion.

To study the magnetic properties of monolayer LiCrTe$_{2}$, it is necessary to construct a spin Hamiltonian, which involves all the dominant magnetic interactions. 
After constructing the spin model with SACE method\cite{LXY,Novel3, NovelNiI2,XCS} and fitting the spin Hamiltonian with MLMCH approach\cite{LXY,Novel2}, the final spin model reads: 

\begin{equation}
    H = H_{2-\textup{order}} + H_{4-\textup{order}}
\end{equation}
where $H_{2-\textup{order}}$ and $H_{4-\textup{order}}$ represent the second (2nd)-order and fourth(4th)-order spin interaction terms, respectively. After collecting the important terms, $H_{2-\textup{order}}$ is found to be expressed as follows: 
\begin{equation}
\begin{aligned}
  &  H_{2-\textup{order}} = \sum_{\langle i,j \rangle_{1}}\{ J_{1} \textbf{S}_{i} \cdot \textbf{S}_{j} + K  S^{\gamma}_{i} S^{\gamma}_{j} + \textbf{D}_{ij} \cdot (\textbf{S}_{i} \times \textbf{S}_{j}) \} \\
 & +  \sum_{\langle i,j \rangle_{2} }J_{2}\textbf{S}_{i} \cdot \textbf{S}_{j} +  \sum_{\langle i,j \rangle_{3}}J_{3}\textbf{S}_{i} \cdot \textbf{S}_{j}  +\sum_{i}A_{zz}(S_{iz})^{2} 
    \end{aligned}
\end{equation} 
with $\langle i,j \rangle_{n}$ ($n = 1,2,3$)  denotes the n-th nearest neighboring spin pairs. $S_{\gamma}$ is the
 $\gamma$ component of the spin vector in a local \{$\alpha \beta \gamma$\} basis (see Fig. S2 in SM\cite{SM}). $J_{i}$ (i = 1,2,3) denotes the isotropic Heisenberg exchange interactions, $K$ denotes the bond-dependent Kitaev interaction, $A_{zz}$ denotes the single-ion anisotropic (SIA), and $\textbf{D}_{ij}$ is the DMI. Considering the $C_{3v}$ symmetry of our system and based on Moriya’s rule, $\textbf{D}_{ij}$ can be further expressed as 
 $d_{//}\hat{\textbf{z}}\times\hat{\textbf{u}}_{ij} + d_{z}\hat{\textbf{z}}$
 , where $\hat{\textbf{u}}_{ij}$ connects two first-nearest-neighbor (1NN) Cr atoms, while $d_{//}$ and $d_{z}$ are the in-plane and out-of-plane components of $\textbf{D}_{ij}$, respectively.

\begin{table}[ht]
\caption{Magnetic parameters with significant values. For simplification, we normalized the spins to $S$ = 1 in this work. The energy unit is meV. One can refers to Eq. (6) and (7) for the exact forms of interactions, and see details in Sec. C, Fig. S4 and Table S2
in SM\cite{SM}.}
\begin{ruledtabular}
    \begin{tabular}{ccccccccc}
      $A_{zz}$ & $J_{1}$ & $K$  & $d_{//}$ & $d_{z}$ & $J_{2}$ & $J_{3}$ &  $B$ & $Y$ \\
    \hline
     -0.05 & -25.53 & 2.30 & -4.74 & 2.01 & -0.43 & 3.95 & -1.96 & -0.35 \\
    \end{tabular}
\end{ruledtabular}
    \label{tb1}
\end{table}

\begin{figure*}[ht]
\includegraphics[scale = 0.31 ]{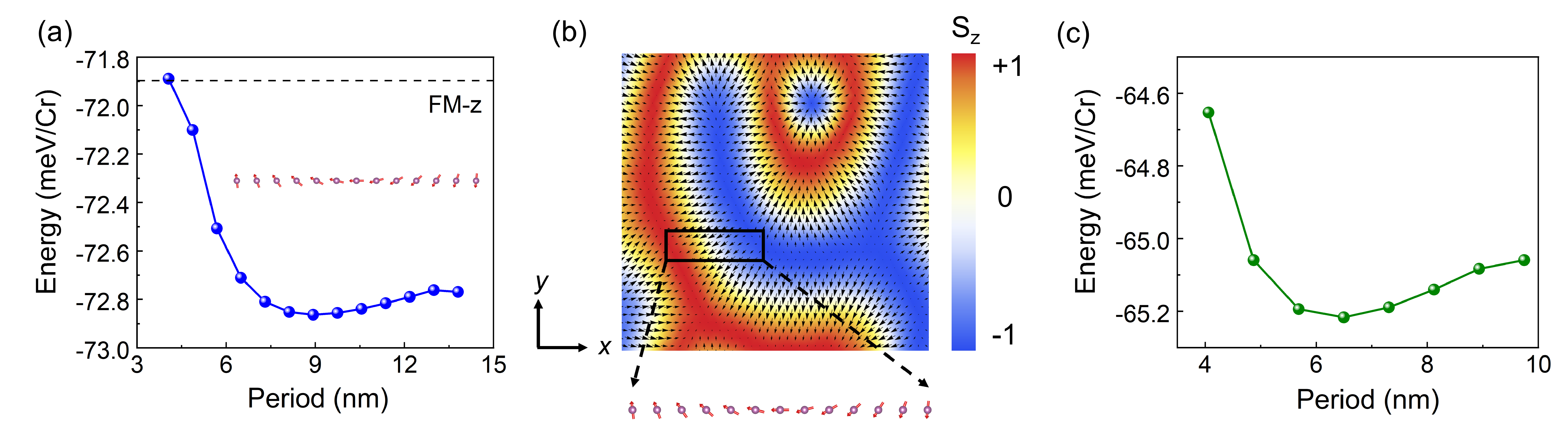}
\caption{\label{2} Magnetic properties of monolayer LiCrTe$_{2}$. (a) The energy of the spin spiral as a function of its period. 
The magnetic configuration with the lowest-energy has a period of 8.93 nm. The spin spirals are illustrated. (b) N$\Acute{\textup{e}}$el-like labyrinth domains (LD). The side view of the selected  spin spirals circled with dashed lines is shown. (c) The energy of the spin spiral as a function of its period but with the higher-order interactions turned off. The magnetic configuration with the lowest-energy has a period of 6.50 nm.
}
\end{figure*}

 As shown in Table \ref{tb1}, the 1NN isotropic Heisenberg exchange yields strong FM coupling with $J_{1}$ = -25.53 meV, and the second-nearest-neighbor (2NN) Heisenberg coupling strength $J_{2}$ = -0.43 meV is rather weak. However, $J_{3}$ = 3.95 meV indicates that the third-nearest-neighbor (3NN) Heisenberg coupling favors antiferromagnetic (AFM) interactions. These two types of interactions compete with each other and give rise to exchange frustration. Moreover, a sizable DMI of $D$ = $\mid \textbf{D}_{ij}\mid $ = 5.15 meV, with $d_{//}$= -4.74 meV and $d_{z}$ = 2.01 meV, is predicted. The resulting $\mid D/J_{1}\mid $ ratio of 0.21, which is quite strong if recalling the typical value of 0.1 $\sim $ 0.2 in skyrmionic system, favors the emergence of chiral spin textures in our system.

 Now, let us turn to the higher-order terms $H_{4-\textup{order}}$. The $H_{4-\textup{order}}$ terms is determined to be:
\begin{equation}
\begin{aligned}
 &  H_{4-\textup{order}} = \sum_{\langle i,j \rangle_{1}}B(\textbf{S}_{i} \cdot \textbf{S}_{j})^{2}  \\
 & + \sum_{\langle i,j,k \rangle}Y \{(\textbf{S}_{i} \cdot \textbf{S}_{j})(\textbf{S}_{j} \cdot \textbf{S}_{k}) + (\textbf{S}_{j} \cdot \textbf{S}_{k})(\textbf{S}_{k} \cdot \textbf{S}_{i})  \\
& + (\textbf{S}_{k} \cdot \textbf{S}_{i})(\textbf{S}_{i} \cdot \textbf{S}_{j})\}+ H_{4-\textup{sites}}  
\end{aligned}
\end{equation}
where $\langle i,j,k \rangle$ denotes the nearest neighboring three-site clusters (see Fig. \ref{1}(d)). $B$ denotes biquadratic interaction, $Y$ denotes isotropic three-site interaction. As shown in Table \ref{tb1}, a strong biquadratic interaction $B$ = -1.96 meV and a smaller but nonneglectable three-site interaction $Y$ = -0.35 meV are predicted. The remaining 4-site interactions $H_{4-\textup{sites}}$, in which its full expression can be found in the Table S2 in SM\cite{SM}, all have negligible magnitudes and are thus not included here. 

To validate the reliability of this model, we compare the energies of different spin configurations predicted by this model with corresponding energies achieved from DFT calculations (see Fig. S3 in SM \cite{SM}) and find that the fitting error is as small as 0.21 meV/Cr. This means that the results predicted from our spin model perfectly match those from DFT calculations. By further including the SOC effect on higher-order interactions, the fitting error would slightly decrease to 0.18 meV/Cr, yet the interaction forms would be complicated and bring huge computational cost. Thus the SOC effect on higher-order couplings, which deserves future investigation, is excluded in our study.

\subsection{Chiral magnetic state: The effects of DMI and fourth-order interactions}
\begin{figure*}[ht]
\includegraphics[scale = 0.31 ]{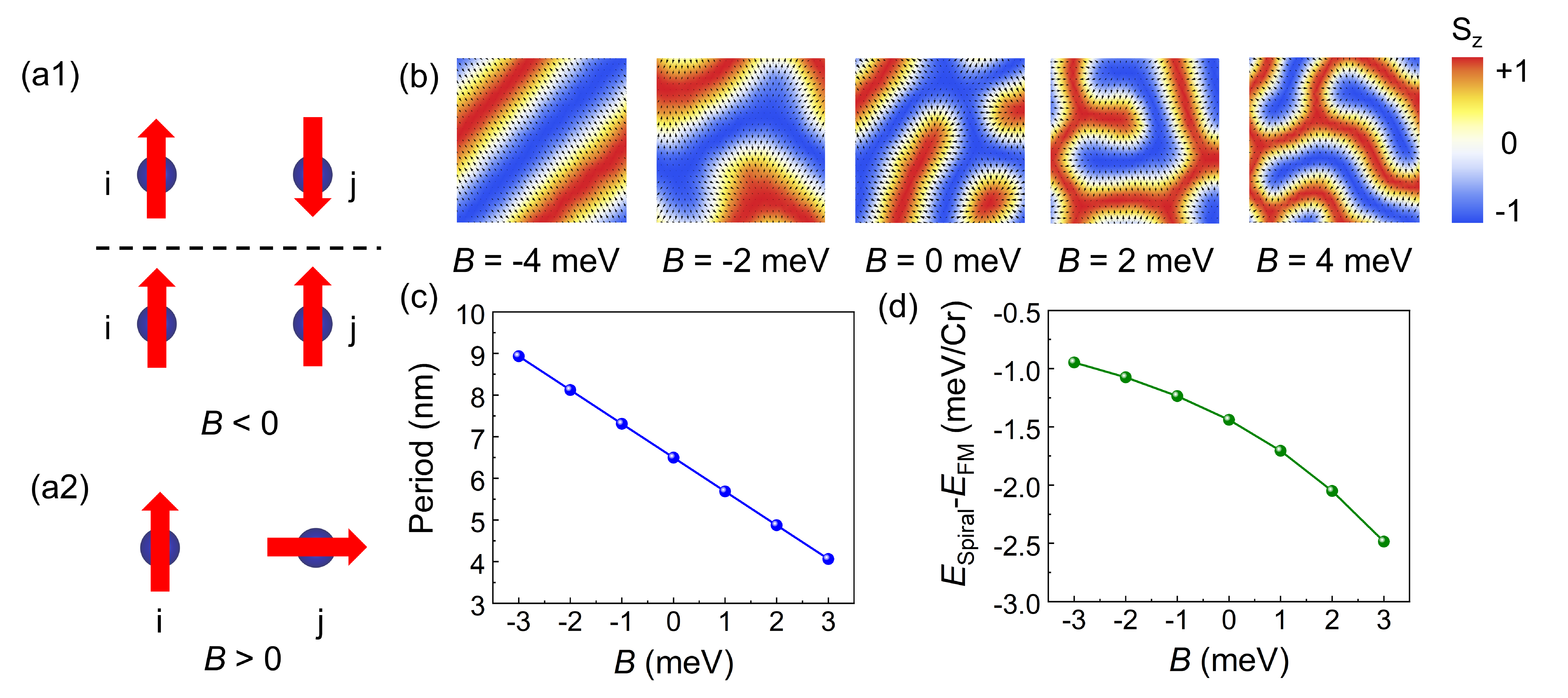}
\caption{\label{3} Role of biquadratic interactions ($B$) in monolayer LiCrTe$_{2}$. (a) The magnetic ground state on a single two-site cluster when only the biquadratic term is turned on. The (b) macroscopic magnetic textures, (c) lowest-energy period of the SS state, and (d) the energy difference between the SS state and FM$_{z}$ state with distinct magnitudes of biquadratic interactions are also shown. Note that in this case, the second-order interactions are all turned on, while the other higher-order interactions besides $B$ are all turned off.
}
\end{figure*}
Based on the aforementioned spin Hamiltonian, MC simulations were carried out to explore the low-energy spin textures of monolayer LiCrTe$_{2}$. The magnetic ground state is identified as a N$\Acute{\textup{e}}$el-type  SS. By further performing MC simulations on a large supercell, one can see that metastable wormlike chiral labyrinth domains (LD) with local topological defects emerge, as shown in Fig. \ref{2}(b). These metastable LD states are assembled by the aforementioned SSs. Meanwhile, one can identify the period of ground state SS state to be 8.93 nm based on the energy of SS as a function of its period(see Fig. \ref{2}(a)). In the following text, we denote this lowest-energy SS state obtained from the spin model with higher-order interactions as the SS$_{1}$ state. This SS$_{1}$ state has a lower energy of -0.97 meV/Cr with respect to the out-of-plane FM (FM$_{z}$) state. 

To understand the microscopic origin of SS$_{1}$ state, we decompose the energy difference between the SS$_{1}$ and FM$_{z}$ states into interaction terms (see Table S1 in SM\cite{SM}). One can see that the largest energy gain of the SS$_{1}$ state with respect to the FM$_{z}$ state is contributed by the in-plane component of the DMI (d$_{//}$), which is responsible for the spin whirling in the x-z plane. Notably. the SS$_{1}$ state would collapse into FM$_{z}$ state without such d$_{//}$ term. This fact demonstrates that d$_{//}$ plays indispensable role on the formation of SS state rather than collinear FM state. Moreover, the d$_{//}$ also chooses the chirality of the SS state. To illustrate this, we artificially construct a SS state, which has the same period with the SS$_{1}$ state by opposite chirality (denoted as SS$^{\textup{OC}}_{1}$). By decomposing the energy difference between SS$_{1}$ state and SS$^{\textup{OC}}_{1}$ state(see Table S1 in SM \cite{SM}), we find that the energy difference is only contributed by d$_{//}$. Furthermore, if one run MC simulations with the value of d$_{//}$ being inverted, a SS with opposite chirality would emerge. From the above results, we thus conclude that it is d$_{//}$ which plays the decisive role on the formation of  SS ground state with certain chirality. Notably, besides d$_{//}$, the AFM $J_{3}$ , which competes with FM $J_{1}$, is also identified to contribute to the energy gain of SS$_{1}$ state with respect to FM$_{z}$ state. This is in line with the fact that exchange frustration between FM 1NN interaction and far-neighboring AFM interactions would help stabilizing chiral spin textures\cite{fru3}.

  \begin{table}[ht]
  \caption{Total energy and relative energies of SS$_{1}$ and SS$_{2}$ states, as well as the decomposition of these energies of some selected critical spin interactions. Note that the 2nd-order (4th-order) means the sum of energy contributions from all 2nd-order (4th-order) spin interactions. The energy contribution of higher-order interactions is shown separately. $B$ and $Y$ means the energy contributions from biquadratic and three-site interactions. 4-sites means the sum of energy contributions from all 4-site spin interactions (No.13, No.14, No.15 in Table S2\cite{SM}). 
  The full energy contributions from all spin interactions is listed in Table S1, Supporting Information\cite{SM}. The energy unit is meV. The value of S is set to be 1. }
  \label{tb2}
  \begin{ruledtabular}
  \begin{tabular}{cccc}
    Coefficient  & SS$_{1}$ &
    SS$_{2}$ & SS$_{1}$ - SS$_{2}$ \\  
    \hline
  Total
 & 
 -72.86
 &
 -72.71
 & -0.15
 \\
  2nd-order   & 
    -65.11 & -65.24   &  
0.13 \\
 4th-order   & 
    -7.75 & -7.47   &  
-0.28 \\
\hline
$B$  & -5.63 &
    -5.43  & -0.20 
\\
    $Y$
    &
    -1.99 & -1.92
    & -0.07
\\
     4-sites
    &
    -0.13 & -0.12
    &
   -0.01 
\\
  \end{tabular}
  \end{ruledtabular}
\end{table}

Interestingly, all the higher-order spin interactions cost energy to the SS$_{1}$ state with respect to the FM$_{z}$ state. This indicates that higher-order interactions could decrease the stability of the SS state, possibly increasing the sensitivity of the SS state under an external field. To identify the effect of higher-order spin interactions on the formation of SS, we perform MC simulations with only 2nd-order interactions, while the higher-order interactions were all turned off. The resulting magnetic ground state without higher-order interactions is still a N$\Acute{\textup{e}}$el-type SS, which is termed the SS$_{2}$ state here. This implies that the higher-order interactions considered in our system would not qualitatively alter the morphology of the magnetic ground state. However, based on the energy of SS as a function of its period (see Fig. \ref{2}(c)), we observed that the period of the SS$_{2}$ state is 6.50 nm, which is remarkably shorter than that of SS$_{1}$ (8.93 nm). This shows that higher-order interactions alter the period of SSs. To gain quantitative insight, we calculated the energies per Cr atom in the SS$_{1}$ and SS$_{2}$ states in our initial spin model with higher-order interactions. As shown in Table \ref{tb2}, the SS$_{2}$ state is 0.15 meV/Cr higher in energy than SS$_{1}$ state. By further decomposing the energy difference between SS$_{1}$ and SS$_{2}$ onto distinct magnetic parameters, one can see that (a) the 2nd-order interactions cause the energy of the SS$_{1}$ state to decrease by 0.13 meV/Cr compared with that of the SS$_{2}$ state, while (b) the higher-order interactions, which are dominantly contributed by biquadratic interactions and three-site interactions, cause the energy of the SS$_{1}$ state to decrease by 0.28 meV/Cr compared with that of the SS$_{2}$ state. From the above results, we can thus conclude that by interplaying with second-order interactions, the higher-order interactions would prefer the emergence of SS state with longer period. 

\begin{figure*}[ht]
\includegraphics[scale = 0.31 ]{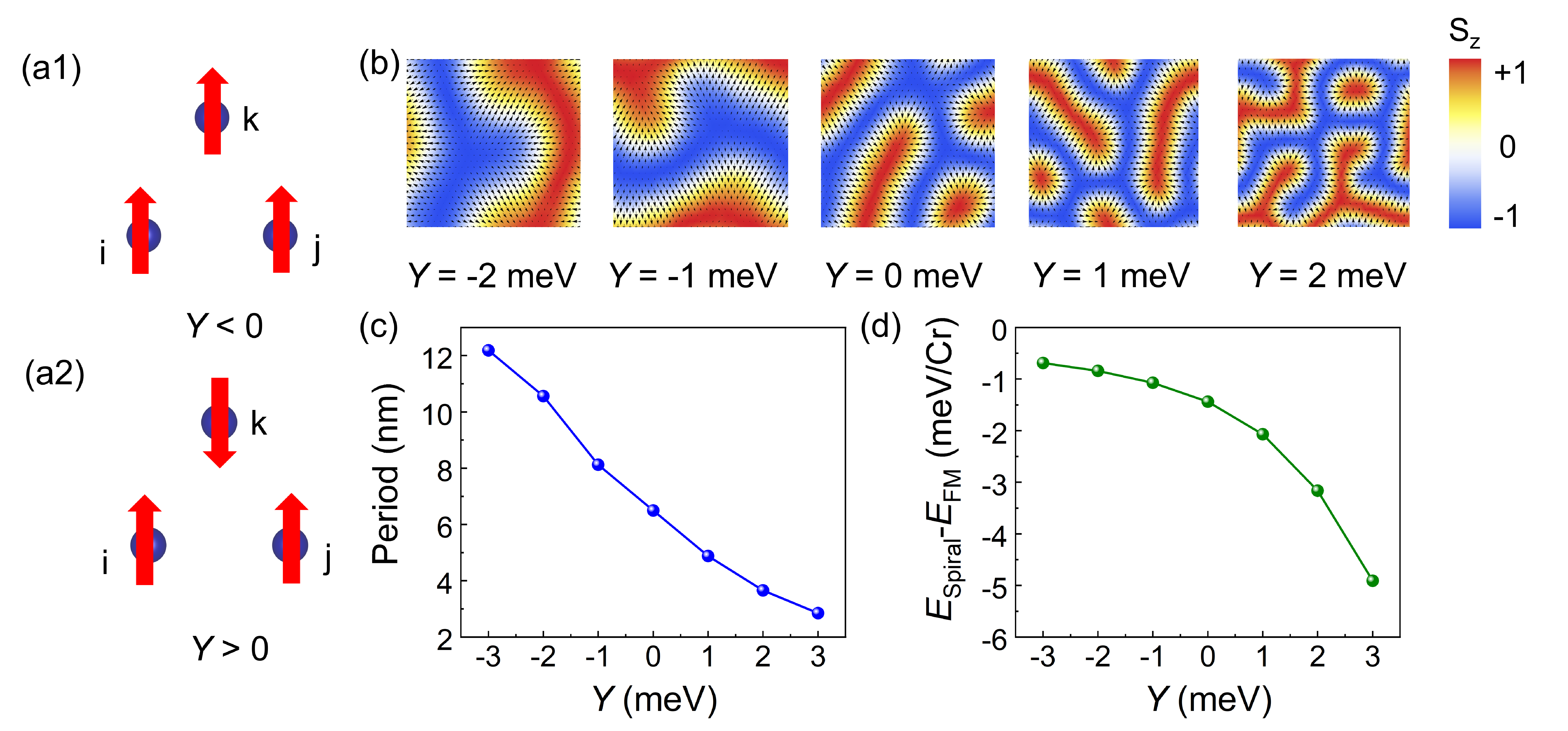}
\caption{\label{4} Role of three-site interactions ($Y$) in monolayer LiCrTe$_{2}$. (a) The magnetic ground state on a single three-site cluster when only the three-site term is turned on. The (b) macroscopic magnetic textures, (c) lowest-energy period of the SS state, and (d) the energy difference between the SS state and FM$_{z}$ state with distinct magnitudes of three-site interactions are also shown. Note that in this case, the second-order interactions are all turned on, while the other higher-order interactions besides $Y$ are all turned off.
}
\end{figure*}

Now, we investigate exactly how biquadratic interactions ($B$) and three-site interactions ($Y$) give rise to the SS configuration with a longer period. First, we focus on the effect of $B$. By acting $B$ on only one nearest-neighboring two-site cluster, one can find that when $B \textless 0$, the nearest-neighboring spins favor a collinear arrangement (up-up or up-down), as shown in Fig. \ref{3}(a1). Such a negative $B$ might suppress the frustration and give rise to a more collinear magnetic state. In contrast, when $B   \textgreater
 0$, the nearest-neighboring spins favor a perpendicular arrangement, as shown in Fig. \ref{3}(a2), which is similar to the situation predicted only by the DMI. Such a positive $B$ could compete with $J_{1}$ and thus lead to noncollinear magnetic textures in the system.

To go further, we attempt to determine how $B$ affects macroscopic spin textures. To this end, we run MC simulations on large supercells with distinct values of $B$, during which 
the 2nd-order interactions are maintained. Meanwhile, the other higher-order interactions are all turned off so that we can focus on the effect of $B$. As shown in Fig. \ref{3}(b), by increasing $B$ from negative to positive values, the macroscopic LDs, which are assembled by SSs, become thinner. To gain quantitative insight, we obtain the lowest-energy SSs under distinct values of $B$, and the predicted period of SSs as a function of $B$ is shown in Fig. \ref{3}(c). When the value of $B$ increases from negative to positive, the period of SSs decreases, which corresponds to the thinner and thinner LDs shown in Fig. \ref{3}(b). Moreover, the energy difference between the FM$_{z}$ state and the SS state as a function of the  $B$ value is shown in Fig. \ref{3}(d). Clearly, by enhancing  $B$ from negative to positive, the energy of SS becomes much lower than that of the FM$_{z}$ state, which indicates that SSs are more stable with positive  $B$. Combining the results above with the fact that  $B$ = -1.96 meV/Cr in our initial model, we can conclude that such negative  $B$ in our system could not only enlarge the period of SSs but also make the SSs more energetically unstable and thus increase the tunability of SSs under external stimulations.

Next, we turn to clarify the effects of $Y$. Similar to the case of $B$ above, we apply $Y$ to only one nearest-neighbor three-site cluster to investigate the preferred magnetic ground state caused by $Y$. As shown in Fig. \ref{4}(a1), when $Y \textless 0$, a ferromagnetic state is preferred for the three-site cluster. In this case, the negative $Y$ could enhance the effect of FM Heisenberg interactions $J_{1}$ acting on nearest-neighboring spins; thus, collinear magnetic textures with less frustration are favored. In contrast, as shown in Fig. \ref{4}(a2), a two-up-one-down AFM spin configuration is favored when $Y \textgreater 0$ on the three-site cluster. In this case, the positive $Y$ could compete with FM $J_{1}$, and the parallel alignment would be weakened, which is benefits to the occurrence of noncollinear spin textures.

\begin{figure*}[ht]
\includegraphics[scale = 0.34 ]{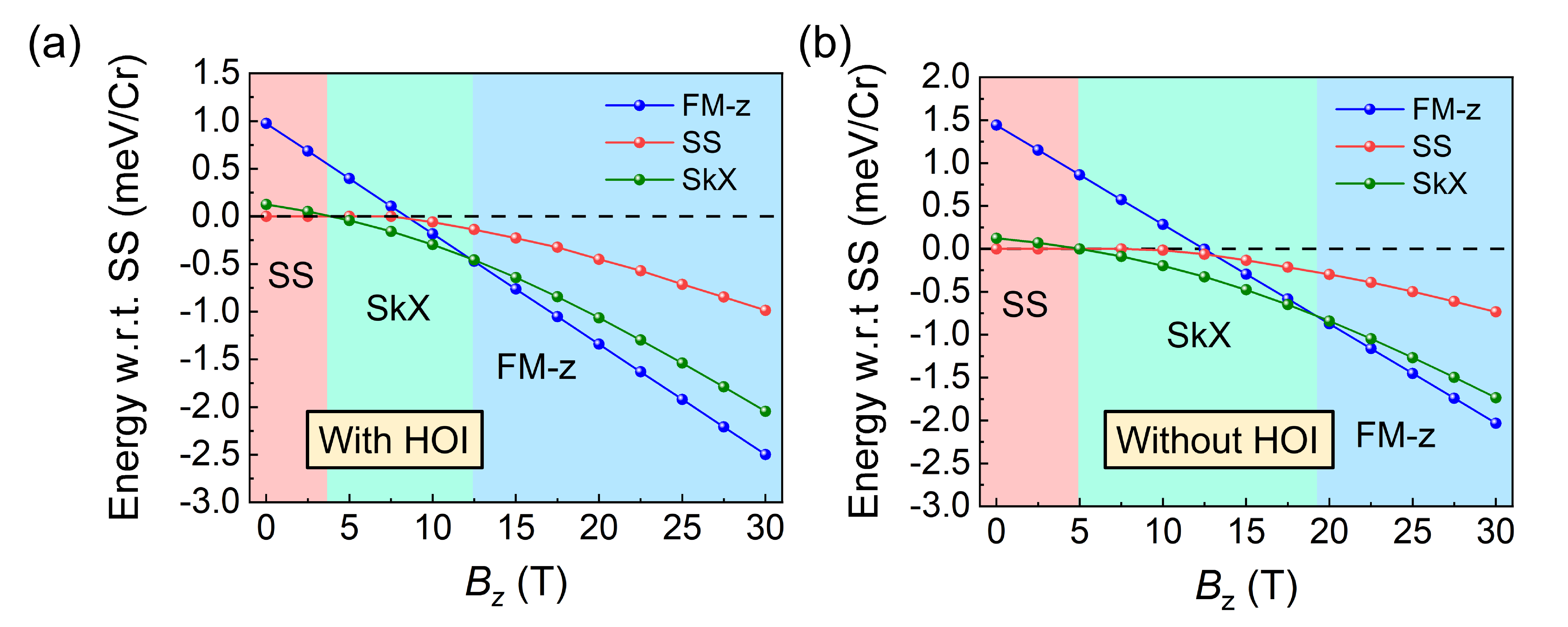}
\caption{\label{5}  Low-temperature phase diagram of monolayer LiCrTe$_{2}$ obtained by (a) including and (b) excluding higher-order interactions. Note that the energies are shown with respect to the zero-field spin spiral state (black dashed line). Here "HOI" means higher-order interactions. 
}
\end{figure*}

We also perform MC simulations on a supercell with distinct $Y$ values to examine the effect of $Y$ on the macroscopic spin textures. Similar with the case of $B$, during our MC simulations, the other higher-order interactions are all turned off, and the 2nd-order interactions are maintained. Such treatment can effectively show the effect of $Y$ and avoid influence from other higher-order interactions. As shown in Fig. \ref{4}(b), by varying $Y$ from negative to positive values, the macroscopic LDs, which are assembled by N$\Acute{\textup{e}}$el-type SS, become thinner. To gain quantitative insight, we constructed the lowest-energy SSs under distinct values of $Y$ and show the period of SSs as a function of $Y$ in Fig. \ref{4}(c). By changing the value of $Y$ from negative to positive, the period of SSs decreases, which qualitatively fits with the thinner LDs shown in Fig. \ref{4}(b). The energy difference between the FM$_{z}$ state and the SS state is also calculated as a function of $Y$, which is shown in Fig. \ref{4}(d). Strikingly, by changing $Y$ from negative to positive values, the SS monotonically lowers its energy and thus becomes much more stable than the FM$_{z}$ state. Recalling that $Y$ = -0.35 meV/Cr, which is a negative value, in our initial model, It is thus clear that such negative $Y$ in our system could increase the thickness of the SSs and the metastable LDs. Moreover, the stability of SSs with respect to the FM$_{z}$ state is suppressed by a positive $Y$, which makes the SSs and the assembled macroscopic LDs more tunable under external fields.

\subsection{Effect of external field on magnetic states}
Since chiral spin textures on a FM background are commonly sensitive to an external magnetic field, the evolution of magnetic states with a magnetic field in our system is naturally considered. To this end, we construct SSs and skyrmion lattices (SkX) with the energetically most favorable sizes and densities. Then, we perform MC simulations with distinct external magnetic fields along the z direction (denoted as $B_{z}$) at a low temperature of 2 K so that the spin structures relax to reach their local equilibrium states. The resulting low-temperature phase diagrams, which show the energies of relaxed SSs, SkX, and FM$_{z}$ states under distinct $B_{z}$, are given in Fig. \ref{5}(a). The SS state has the lowest energy without an external magnetic field. However, at a certain critical field value of approximately 3 T, SkX is energetically favorable than SS state. For a larger critical value of approximately 12.5 T, SkX is suppressed, and the system prefers the FM$_{z}$ state.

As previously noted, higher-order interactions decrease the energy difference between FM$_{z}$ and the SS state, which might affect the stability of chiral spin textures under a magnetic field. Here, we provide quantitative insight into this phenomenon by calculating a low-temperature phase diagram without the inclusion of any higher-order interactions. As shown in Fig. \ref{5}(b), the magnetic phase transition under $B_{z}$ (SS is evaluated into SkX and then into FM$_{z}$) is similar to that in the presence of higher-order interactions. However, one can still find two qualitative aspects: (a) the value of the critical field for SS to evaluate SkX increases from approximately 3.5 T to approximately 5 T, which means that the SS would be more robust after excluding the effect of higher-order interactions; (b) the value of the critical field for SkX to transforming to FM$_{z}$ increases from approximately 12.5 T to approximately 19 T, which implies that a larger magnetic field is needed to fully suppress the chiral magnetic textures without the inclusion of higher-order interactions. These results suggest that the higher-order interactions in our system increase the sensitivity of the chiral magnetic textures to the external field, which is beneficial for realizing field-sensitive spintronics devices in the future.

\section{CONCLUSIONS}
In conclusion, by using the symmetry-adapted cluster expansion method, we constructed a first-principle-based spin model for monolayer LiCrTe$_{2}$, which predicted a chiral spin spiral state as the ground state. Moreover, such model reveals the existence of higher-order interactions in the system, which would expand the size of spin spirals and increase the field tunability of chiral spin textures under external magnetic fields. Our work not only provides comprehensive knowledge on the complex magnetic interactions in 2D magnetic systems but also sheds lights on the field-tunable spintronics devices based on chiral spin textures.

\section{Acknowledgments}
We thank Xuan Zou and Binhua Zhang for helpful discussion. C.X.  acknowledges financial support from the Ministry of Science and Technology of the People's Republic of China (No. 2022YFA1402901), NSFC (grants No.  12274082), Shanghai Science and Technology Committee (grant No. 23ZR1406600) and the open project of Guangdong provincial key laboratory of magnetoelectric physics and devices (No. 2022B1212010008). W.D. acknowledges financial support from the Basic Science Center Project of NSFC and the Beijing Advanced Innovation Center for Future Chip (ICFC). 
\bibliographystyle{apsrev4-1}

\begin{thebibliography}{79}%
\makeatletter
\providecommand \@ifxundefined [1]{%
 \@ifx{#1\undefined}
}%
\providecommand \@ifnum [1]{%
 \ifnum #1\expandafter \@firstoftwo
 \else \expandafter \@secondoftwo
 \fi
}%
\providecommand \@ifx [1]{%
 \ifx #1\expandafter \@firstoftwo
 \else \expandafter \@secondoftwo
 \fi
}%
\providecommand \natexlab [1]{#1}%
\providecommand \enquote  [1]{``#1''}%
\providecommand \bibnamefont  [1]{#1}%
\providecommand \bibfnamefont [1]{#1}%
\providecommand \citenamefont [1]{#1}%
\providecommand \href@noop [0]{\@secondoftwo}%
\providecommand \href [0]{\begingroup \@sanitize@url \@href}%
\providecommand \@href[1]{\@@startlink{#1}\@@href}%
\providecommand \@@href[1]{\endgroup#1\@@endlink}%
\providecommand \@sanitize@url [0]{\catcode `\\12\catcode `\$12\catcode
  `\&12\catcode `\#12\catcode `\^12\catcode `\_12\catcode `\%12\relax}%
\providecommand \@@startlink[1]{}%
\providecommand \@@endlink[0]{}%
\providecommand \url  [0]{\begingroup\@sanitize@url \@url }%
\providecommand \@url [1]{\endgroup\@href {#1}{\urlprefix }}%
\providecommand \urlprefix  [0]{URL }%
\providecommand \Eprint [0]{\href }%
\providecommand \doibase [0]{http://dx.doi.org/}%
\providecommand \selectlanguage [0]{\@gobble}%
\providecommand \bibinfo  [0]{\@secondoftwo}%
\providecommand \bibfield  [0]{\@secondoftwo}%
\providecommand \translation [1]{[#1]}%
\providecommand \BibitemOpen [0]{}%
\providecommand \bibitemStop [0]{}%
\providecommand \bibitemNoStop [0]{.\EOS\space}%
\providecommand \EOS [0]{\spacefactor3000\relax}%
\providecommand \BibitemShut  [1]{\csname bibitem#1\endcsname}%
\let\auto@bib@innerbib\@empty
\bibitem [{\citenamefont {Bode}\ \emph {et~al.}(2007)\citenamefont {Bode},
  \citenamefont {Heide}, \citenamefont {Von~Bergmann}, \citenamefont
  {Ferriani}, \citenamefont {Heinze}, \citenamefont {Bihlmayer}, \citenamefont
  {Kubetzka}, \citenamefont {Pietzsch}, \citenamefont {Bl{\"u}gel},\ and\
  \citenamefont {Wiesendanger}}]{intro1}%
  \BibitemOpen
  \bibfield  {author} {\bibinfo {author} {\bibfnamefont {M.}~\bibnamefont
  {Bode}}, \bibinfo {author} {\bibfnamefont {M.}~\bibnamefont {Heide}},
  \bibinfo {author} {\bibfnamefont {K.}~\bibnamefont {Von~Bergmann}}, \bibinfo
  {author} {\bibfnamefont {P.}~\bibnamefont {Ferriani}}, \bibinfo {author}
  {\bibfnamefont {S.}~\bibnamefont {Heinze}}, \bibinfo {author} {\bibfnamefont
  {G.}~\bibnamefont {Bihlmayer}}, \bibinfo {author} {\bibfnamefont
  {A.}~\bibnamefont {Kubetzka}}, \bibinfo {author} {\bibfnamefont
  {O.}~\bibnamefont {Pietzsch}}, \bibinfo {author} {\bibfnamefont
  {S.}~\bibnamefont {Bl{\"u}gel}}, \ and\ \bibinfo {author} {\bibfnamefont
  {R.}~\bibnamefont {Wiesendanger}},\ }\href
  {https://www.nature.com/articles/nature05802} {\bibfield  {journal} {\bibinfo
   {journal} {Nature}\ }\textbf {\bibinfo {volume} {447}},\ \bibinfo {pages}
  {190} (\bibinfo {year} {2007})}\BibitemShut {NoStop}%
\bibitem [{\citenamefont {Franken}\ \emph {et~al.}(2014)\citenamefont
  {Franken}, \citenamefont {Herps}, \citenamefont {Swagten},\ and\
  \citenamefont {Koopmans}}]{intro2}%
  \BibitemOpen
  \bibfield  {author} {\bibinfo {author} {\bibfnamefont {J.~H.}\ \bibnamefont
  {Franken}}, \bibinfo {author} {\bibfnamefont {M.}~\bibnamefont {Herps}},
  \bibinfo {author} {\bibfnamefont {H.~J.}\ \bibnamefont {Swagten}}, \ and\
  \bibinfo {author} {\bibfnamefont {B.}~\bibnamefont {Koopmans}},\ }\href
  {https://www.nature.com/articles/srep05248} {\bibfield  {journal} {\bibinfo
  {journal} {Sci. Rep.}\ }\textbf {\bibinfo {volume} {4}},\ \bibinfo {pages}
  {1} (\bibinfo {year} {2014})}\BibitemShut {NoStop}%
\bibitem [{\citenamefont {Ryu}\ \emph {et~al.}(2013)\citenamefont {Ryu},
  \citenamefont {Thomas}, \citenamefont {Yang},\ and\ \citenamefont
  {Parkin}}]{intro3}%
  \BibitemOpen
  \bibfield  {author} {\bibinfo {author} {\bibfnamefont {K.-S.}\ \bibnamefont
  {Ryu}}, \bibinfo {author} {\bibfnamefont {L.}~\bibnamefont {Thomas}},
  \bibinfo {author} {\bibfnamefont {S.-H.}\ \bibnamefont {Yang}}, \ and\
  \bibinfo {author} {\bibfnamefont {S.}~\bibnamefont {Parkin}},\ }\href
  {https://www.nature.com/articles/nnano.2013.102} {\bibfield  {journal}
  {\bibinfo  {journal} {Nat. Nanotechnol.}\ }\textbf {\bibinfo {volume} {8}},\
  \bibinfo {pages} {527} (\bibinfo {year} {2013})}\BibitemShut {NoStop}%
\bibitem [{\citenamefont {Mühlbauer}\ \emph {et~al.}(2009)\citenamefont
  {Mühlbauer}, \citenamefont {Binz}, \citenamefont {Jonietz}, \citenamefont
  {Pfleiderer}, \citenamefont {Rosch}, \citenamefont {Neubauer}, \citenamefont
  {Georgii},\ and\ \citenamefont {B\"oni}}]{skyrmion1}%
  \BibitemOpen
  \bibfield  {author} {\bibinfo {author} {\bibfnamefont {S.}~\bibnamefont
  {Muhlba\"uer}}, \bibinfo {author} {\bibfnamefont {B.}~\bibnamefont {Binz}},
  \bibinfo {author} {\bibfnamefont {F.}~\bibnamefont {Jonietz}}, \bibinfo
  {author} {\bibfnamefont {C.}~\bibnamefont {Pfleiderer}}, \bibinfo {author}
  {\bibfnamefont {A.}~\bibnamefont {Rosch}}, \bibinfo {author} {\bibfnamefont
  {A.}~\bibnamefont {Neubauer}}, \bibinfo {author} {\bibfnamefont
  {R.}~\bibnamefont {Georgii}}, \ and\ \bibinfo {author} {\bibfnamefont
  {P.}~\bibnamefont {B\"oni}},\ }\href
  {https://www.science.org/doi/full/10.1126/science.1166767} {\bibfield
  {journal} {\bibinfo  {journal} {Science}\ }\textbf {\bibinfo {volume}
  {323}},\ \bibinfo {pages} {915} (\bibinfo {year} {2009})}\BibitemShut
  {NoStop}%
\bibitem [{\citenamefont {Yu}\ \emph {et~al.}(2010)\citenamefont {Yu},
  \citenamefont {Onose}, \citenamefont {Kanazawa}, \citenamefont {Park},
  \citenamefont {Han}, \citenamefont {Matsui}, \citenamefont {Nagaosa},\ and\
  \citenamefont {Tokura}}]{skyrmion2}%
  \BibitemOpen
  \bibfield  {author} {\bibinfo {author} {\bibfnamefont {X.}~\bibnamefont
  {Yu}}, \bibinfo {author} {\bibfnamefont {Y.}~\bibnamefont {Onose}}, \bibinfo
  {author} {\bibfnamefont {N.}~\bibnamefont {Kanazawa}}, \bibinfo {author}
  {\bibfnamefont {J.~H.}\ \bibnamefont {Park}}, \bibinfo {author}
  {\bibfnamefont {J.}~\bibnamefont {Han}}, \bibinfo {author} {\bibfnamefont
  {Y.}~\bibnamefont {Matsui}}, \bibinfo {author} {\bibfnamefont
  {N.}~\bibnamefont {Nagaosa}}, \ and\ \bibinfo {author} {\bibfnamefont
  {Y.}~\bibnamefont {Tokura}},\ }\href
  {https://www.nature.com/articles/nature09124} {\bibfield  {journal} {\bibinfo
   {journal} {Nature}\ }\textbf {\bibinfo {volume} {465}},\ \bibinfo {pages}
  {901} (\bibinfo {year} {2010})}\BibitemShut {NoStop}%
\bibitem [{\citenamefont {Fert}\ \emph {et~al.}(2017)\citenamefont {Fert},
  \citenamefont {Reyren},\ and\ \citenamefont {Cros}}]{skyrmion3}%
  \BibitemOpen
  \bibfield  {author} {\bibinfo {author} {\bibfnamefont {A.}~\bibnamefont
  {Fert}}, \bibinfo {author} {\bibfnamefont {N.}~\bibnamefont {Reyren}}, \ and\
  \bibinfo {author} {\bibfnamefont {V.}~\bibnamefont {Cros}},\ }\href
  {https://www.nature.com/articles/natrevmats201731} {\bibfield  {journal}
  {\bibinfo  {journal} {Nat. Rev. Mater.}\ }\textbf {\bibinfo {volume} {2}},\
  \bibinfo {pages} {1} (\bibinfo {year} {2017})}\BibitemShut {NoStop}%
\bibitem [{\citenamefont {Bhowal}\ and\ \citenamefont
  {Spaldin}(2022)}]{SkxME1}%
  \BibitemOpen
  \bibfield  {author} {\bibinfo {author} {\bibfnamefont {S.}~\bibnamefont
  {Bhowal}}\ and\ \bibinfo {author} {\bibfnamefont {N.~A.}\ \bibnamefont
  {Spaldin}},\ }\href {\doibase 10.1103/PhysRevLett.128.227204} {\bibfield
  {journal} {\bibinfo  {journal} {Phys. Rev. Lett.}\ }\textbf {\bibinfo
  {volume} {128}},\ \bibinfo {pages} {227204} (\bibinfo {year}
  {2022})}\BibitemShut {NoStop}%
\bibitem [{\citenamefont {Ederer}\ and\ \citenamefont
  {Spaldin}(2007)}]{SkxME2}%
  \BibitemOpen
  \bibfield  {author} {\bibinfo {author} {\bibfnamefont {C.}~\bibnamefont
  {Ederer}}\ and\ \bibinfo {author} {\bibfnamefont {N.~A.}\ \bibnamefont
  {Spaldin}},\ }\href {\doibase 10.1103/PhysRevB.76.214404} {\bibfield
  {journal} {\bibinfo  {journal} {Phys. Rev. B}\ }\textbf {\bibinfo {volume}
  {76}},\ \bibinfo {pages} {214404} (\bibinfo {year} {2007})}\BibitemShut
  {NoStop}%
\bibitem [{\citenamefont {Li}\ \emph {et~al.}(2013)\citenamefont {Li},
  \citenamefont {Kanazawa}, \citenamefont {Yu}, \citenamefont {Tsukazaki},
  \citenamefont {Kawasaki}, \citenamefont {Ichikawa}, \citenamefont {Jin},
  \citenamefont {Kagawa},\ and\ \citenamefont {Tokura}}]{SkxME3}%
  \BibitemOpen
  \bibfield  {author} {\bibinfo {author} {\bibfnamefont {Y.}~\bibnamefont
  {Li}}, \bibinfo {author} {\bibfnamefont {N.}~\bibnamefont {Kanazawa}},
  \bibinfo {author} {\bibfnamefont {X.~Z.}\ \bibnamefont {Yu}}, \bibinfo
  {author} {\bibfnamefont {A.}~\bibnamefont {Tsukazaki}}, \bibinfo {author}
  {\bibfnamefont {M.}~\bibnamefont {Kawasaki}}, \bibinfo {author}
  {\bibfnamefont {M.}~\bibnamefont {Ichikawa}}, \bibinfo {author}
  {\bibfnamefont {X.~F.}\ \bibnamefont {Jin}}, \bibinfo {author} {\bibfnamefont
  {F.}~\bibnamefont {Kagawa}}, \ and\ \bibinfo {author} {\bibfnamefont
  {Y.}~\bibnamefont {Tokura}},\ }\href
  {https://link.aps.org/doi/10.1103/PhysRevLett.110.117202} {\bibfield
  {journal} {\bibinfo  {journal} {Phys. Rev. Lett.}\ }\textbf {\bibinfo
  {volume} {110}},\ \bibinfo {pages} {117202} (\bibinfo {year}
  {2013})}\BibitemShut {NoStop}%
\bibitem [{\citenamefont {Schulz}\ \emph {et~al.}(2012)\citenamefont {Schulz},
  \citenamefont {Ritz}, \citenamefont {Bauer}, \citenamefont {Halder},
  \citenamefont {Wagner}, \citenamefont {Franz}, \citenamefont {Pfleiderer},
  \citenamefont {Everschor}, \citenamefont {Garst},\ and\ \citenamefont
  {Rosch}}]{SkxME4}%
  \BibitemOpen
  \bibfield  {author} {\bibinfo {author} {\bibfnamefont {T.}~\bibnamefont
  {Schulz}}, \bibinfo {author} {\bibfnamefont {R.}~\bibnamefont {Ritz}},
  \bibinfo {author} {\bibfnamefont {A.}~\bibnamefont {Bauer}}, \bibinfo
  {author} {\bibfnamefont {M.}~\bibnamefont {Halder}}, \bibinfo {author}
  {\bibfnamefont {M.}~\bibnamefont {Wagner}}, \bibinfo {author} {\bibfnamefont
  {C.}~\bibnamefont {Franz}}, \bibinfo {author} {\bibfnamefont
  {C.}~\bibnamefont {Pfleiderer}}, \bibinfo {author} {\bibfnamefont
  {K.}~\bibnamefont {Everschor}}, \bibinfo {author} {\bibfnamefont
  {M.}~\bibnamefont {Garst}}, \ and\ \bibinfo {author} {\bibfnamefont
  {A.}~\bibnamefont {Rosch}},\ }\href
  {https://www.nature.com/articles/nphys2231} {\bibfield  {journal} {\bibinfo
  {journal} {Nat. Phys.}\ }\textbf {\bibinfo {volume} {8}},\ \bibinfo {pages}
  {301} (\bibinfo {year} {2012})}\BibitemShut {NoStop}%
\bibitem [{\citenamefont {Yang}\ \emph {et~al.}(2021)\citenamefont {Yang},
  \citenamefont {Naaman}, \citenamefont {Paltiel},\ and\ \citenamefont
  {Parkin}}]{SkXApp4}%
  \BibitemOpen
  \bibfield  {author} {\bibinfo {author} {\bibfnamefont {S.-H.}\ \bibnamefont
  {Yang}}, \bibinfo {author} {\bibfnamefont {R.}~\bibnamefont {Naaman}},
  \bibinfo {author} {\bibfnamefont {Y.}~\bibnamefont {Paltiel}}, \ and\
  \bibinfo {author} {\bibfnamefont {S.~S.}\ \bibnamefont {Parkin}},\ }\href
  {https://www.nature.com/articles/s42254-021-00302-9} {\bibfield  {journal}
  {\bibinfo  {journal} {Nat. Rev. Phys.}\ }\textbf {\bibinfo {volume} {3}},\
  \bibinfo {pages} {328} (\bibinfo {year} {2021})}\BibitemShut {NoStop}%
\bibitem [{\citenamefont {Parkin}\ \emph {et~al.}(2008)\citenamefont {Parkin},
  \citenamefont {Hayashi},\ and\ \citenamefont {Thomas}}]{SkXApp2}%
  \BibitemOpen
  \bibfield  {author} {\bibinfo {author} {\bibfnamefont {S.~S.}\ \bibnamefont
  {Parkin}}, \bibinfo {author} {\bibfnamefont {M.}~\bibnamefont {Hayashi}}, \
  and\ \bibinfo {author} {\bibfnamefont {L.}~\bibnamefont {Thomas}},\ }\href
  {https://www.science.org/doi/full/10.1126/science.1145799} {\bibfield
  {journal} {\bibinfo  {journal} {Science}\ }\textbf {\bibinfo {volume}
  {320}},\ \bibinfo {pages} {190} (\bibinfo {year} {2008})}\BibitemShut
  {NoStop}%
\bibitem [{\citenamefont {Mochizuki}\ and\ \citenamefont
  {Seki}(2013)}]{SkXApp1}%
  \BibitemOpen
  \bibfield  {author} {\bibinfo {author} {\bibfnamefont {M.}~\bibnamefont
  {Mochizuki}}\ and\ \bibinfo {author} {\bibfnamefont {S.}~\bibnamefont
  {Seki}},\ }\href {\doibase 10.1103/PhysRevB.87.134403} {\bibfield  {journal}
  {\bibinfo  {journal} {Phys. Rev. B}\ }\textbf {\bibinfo {volume} {87}},\
  \bibinfo {pages} {134403} (\bibinfo {year} {2013})}\BibitemShut {NoStop}%
\bibitem [{\citenamefont {Haldar}\ \emph {et~al.}(2018)\citenamefont {Haldar},
  \citenamefont {von Malottki}, \citenamefont {Meyer}, \citenamefont
  {Bessarab},\ and\ \citenamefont {Heinze}}]{STB1}%
  \BibitemOpen
  \bibfield  {author} {\bibinfo {author} {\bibfnamefont {S.}~\bibnamefont
  {Haldar}}, \bibinfo {author} {\bibfnamefont {S.}~\bibnamefont {von
  Malottki}}, \bibinfo {author} {\bibfnamefont {S.}~\bibnamefont {Meyer}},
  \bibinfo {author} {\bibfnamefont {P.~F.}\ \bibnamefont {Bessarab}}, \ and\
  \bibinfo {author} {\bibfnamefont {S.}~\bibnamefont {Heinze}},\ }\href
  {\doibase 10.1103/PhysRevB.98.060413} {\bibfield  {journal} {\bibinfo
  {journal} {Phys. Rev. B}\ }\textbf {\bibinfo {volume} {98}},\ \bibinfo
  {pages} {060413(R)} (\bibinfo {year} {2018})}\BibitemShut {NoStop}%
\bibitem [{\citenamefont {Goerzen}\ \emph {et~al.}(2022)\citenamefont
  {Goerzen}, \citenamefont {von Malottki}, \citenamefont {Kwiatkowski},
  \citenamefont {Bessarab},\ and\ \citenamefont {Heinze}}]{STB2}%
  \BibitemOpen
  \bibfield  {author} {\bibinfo {author} {\bibfnamefont {M.~A.}\ \bibnamefont
  {Goerzen}}, \bibinfo {author} {\bibfnamefont {S.}~\bibnamefont {von
  Malottki}}, \bibinfo {author} {\bibfnamefont {G.~J.}\ \bibnamefont
  {Kwiatkowski}}, \bibinfo {author} {\bibfnamefont {P.~F.}\ \bibnamefont
  {Bessarab}}, \ and\ \bibinfo {author} {\bibfnamefont {S.}~\bibnamefont
  {Heinze}},\ }\href {\doibase 10.1103/PhysRevB.105.214435} {\bibfield
  {journal} {\bibinfo  {journal} {Phys. Rev. B}\ }\textbf {\bibinfo {volume}
  {105}},\ \bibinfo {pages} {214435} (\bibinfo {year} {2022})}\BibitemShut
  {NoStop}%
\bibitem [{\citenamefont {Li}\ \emph {et~al.}(2022{\natexlab{a}})\citenamefont
  {Li}, \citenamefont {Haldar},\ and\ \citenamefont {Heinze}}]{STB3}%
  \BibitemOpen
  \bibfield  {author} {\bibinfo {author} {\bibfnamefont {D.}~\bibnamefont
  {Li}}, \bibinfo {author} {\bibfnamefont {S.}~\bibnamefont {Haldar}}, \ and\
  \bibinfo {author} {\bibfnamefont {S.}~\bibnamefont {Heinze}},\ }\href
  {https://pubs.acs.org/doi/abs/10.1021/acs.nanolett.2c03287} {\bibfield
  {journal} {\bibinfo  {journal} {Nano Lett.}\ }\textbf {\bibinfo {volume}
  {22}},\ \bibinfo {pages} {7706} (\bibinfo {year}
  {2022}{\natexlab{a}})}\BibitemShut {NoStop}%
\bibitem [{\citenamefont {Li}\ \emph {et~al.}(2024)\citenamefont {Li},
  \citenamefont {Haldar},\ and\ \citenamefont {Heinze}}]{STB4}%
  \BibitemOpen
  \bibfield  {author} {\bibinfo {author} {\bibfnamefont {D.}~\bibnamefont
  {Li}}, \bibinfo {author} {\bibfnamefont {S.}~\bibnamefont {Haldar}}, \ and\
  \bibinfo {author} {\bibfnamefont {S.}~\bibnamefont {Heinze}},\ }\href
  {https://arxiv.org/abs/2401.18000} {\bibfield  {journal} {\bibinfo  {journal}
  {arXiv:2401.18000}\ } (\bibinfo {year} {2024})}\BibitemShut {NoStop}%
\bibitem [{\citenamefont {Varentcova}\ \emph {et~al.}(2020)\citenamefont
  {Varentcova}, \citenamefont {von Malottki}, \citenamefont {Potkina},
  \citenamefont {Kwiatkowski}, \citenamefont {Heinze},\ and\ \citenamefont
  {Bessarab}}]{STB5}%
  \BibitemOpen
  \bibfield  {author} {\bibinfo {author} {\bibfnamefont {A.~S.}\ \bibnamefont
  {Varentcova}}, \bibinfo {author} {\bibfnamefont {S.}~\bibnamefont {von
  Malottki}}, \bibinfo {author} {\bibfnamefont {M.~N.}\ \bibnamefont
  {Potkina}}, \bibinfo {author} {\bibfnamefont {G.}~\bibnamefont
  {Kwiatkowski}}, \bibinfo {author} {\bibfnamefont {S.}~\bibnamefont {Heinze}},
  \ and\ \bibinfo {author} {\bibfnamefont {P.~F.}\ \bibnamefont {Bessarab}},\
  }\href {https://www.nature.com/articles/s41524-020-00453-w} {\bibfield
  {journal} {\bibinfo  {journal} {npj Comput Mater}\ }\textbf {\bibinfo
  {volume} {6}},\ \bibinfo {pages} {193} (\bibinfo {year} {2020})}\BibitemShut
  {NoStop}%
\bibitem [{\citenamefont {He}\ \emph {et~al.}(2024)\citenamefont {He},
  \citenamefont {Dou}, \citenamefont {Du}, \citenamefont {Dai}, \citenamefont
  {Huang},\ and\ \citenamefont {Ma}}]{DMIType}%
  \BibitemOpen
  \bibfield  {author} {\bibinfo {author} {\bibfnamefont {Z.}~\bibnamefont
  {He}}, \bibinfo {author} {\bibfnamefont {K.}~\bibnamefont {Dou}}, \bibinfo
  {author} {\bibfnamefont {W.}~\bibnamefont {Du}}, \bibinfo {author}
  {\bibfnamefont {Y.}~\bibnamefont {Dai}}, \bibinfo {author} {\bibfnamefont
  {B.}~\bibnamefont {Huang}}, \ and\ \bibinfo {author} {\bibfnamefont
  {Y.}~\bibnamefont {Ma}},\ }\href {\doibase 10.1103/PhysRevB.109.024420}
  {\bibfield  {journal} {\bibinfo  {journal} {Phys. Rev. B}\ }\textbf {\bibinfo
  {volume} {109}},\ \bibinfo {pages} {024420} (\bibinfo {year}
  {2024})}\BibitemShut {NoStop}%
\bibitem [{\citenamefont {Huang}\ \emph {et~al.}(2024)\citenamefont {Huang},
  \citenamefont {Schwartz}, \citenamefont {Shao}, \citenamefont {Kovalev},\
  and\ \citenamefont {Tsymbal}}]{DMIType2}%
  \BibitemOpen
  \bibfield  {author} {\bibinfo {author} {\bibfnamefont {K.}~\bibnamefont
  {Huang}}, \bibinfo {author} {\bibfnamefont {E.}~\bibnamefont {Schwartz}},
  \bibinfo {author} {\bibfnamefont {D.-F.}\ \bibnamefont {Shao}}, \bibinfo
  {author} {\bibfnamefont {A.~A.}\ \bibnamefont {Kovalev}}, \ and\ \bibinfo
  {author} {\bibfnamefont {E.~Y.}\ \bibnamefont {Tsymbal}},\ }\href
  {https://link.aps.org/doi/10.1103/PhysRevB.109.024426} {\bibfield  {journal}
  {\bibinfo  {journal} {Phys. Rev. B}\ }\textbf {\bibinfo {volume} {109}},\
  \bibinfo {pages} {024426} (\bibinfo {year} {2024})}\BibitemShut {NoStop}%
\bibitem [{\citenamefont {Hoffmann}\ \emph {et~al.}(2017)\citenamefont
  {Hoffmann}, \citenamefont {Zimmermann}, \citenamefont {M{\"u}ller},
  \citenamefont {Sch{\"u}rhoff}, \citenamefont {Kiselev}, \citenamefont
  {Melcher},\ and\ \citenamefont {Bl{\"u}gel}}]{DMIType3}%
  \BibitemOpen
  \bibfield  {author} {\bibinfo {author} {\bibfnamefont {M.}~\bibnamefont
  {Hoffmann}}, \bibinfo {author} {\bibfnamefont {B.}~\bibnamefont
  {Zimmermann}}, \bibinfo {author} {\bibfnamefont {G.~P.}\ \bibnamefont
  {M{\"u}ller}}, \bibinfo {author} {\bibfnamefont {D.}~\bibnamefont
  {Sch{\"u}rhoff}}, \bibinfo {author} {\bibfnamefont {N.~S.}\ \bibnamefont
  {Kiselev}}, \bibinfo {author} {\bibfnamefont {C.}~\bibnamefont {Melcher}}, \
  and\ \bibinfo {author} {\bibfnamefont {S.}~\bibnamefont {Bl{\"u}gel}},\
  }\href {https://www.nature.com/articles/s41467-017-00313-0} {\bibfield
  {journal} {\bibinfo  {journal} {Nat Commun}\ }\textbf {\bibinfo {volume}
  {8}},\ \bibinfo {pages} {308} (\bibinfo {year} {2017})}\BibitemShut {NoStop}%
\bibitem [{\citenamefont {Du}\ \emph {et~al.}(2023)\citenamefont {Du},
  \citenamefont {Dou}, \citenamefont {He}, \citenamefont {Dai}, \citenamefont
  {Huang},\ and\ \citenamefont {Ma}}]{DMIType4}%
  \BibitemOpen
  \bibfield  {author} {\bibinfo {author} {\bibfnamefont {W.}~\bibnamefont
  {Du}}, \bibinfo {author} {\bibfnamefont {K.}~\bibnamefont {Dou}}, \bibinfo
  {author} {\bibfnamefont {Z.}~\bibnamefont {He}}, \bibinfo {author}
  {\bibfnamefont {Y.}~\bibnamefont {Dai}}, \bibinfo {author} {\bibfnamefont
  {B.}~\bibnamefont {Huang}}, \ and\ \bibinfo {author} {\bibfnamefont
  {Y.}~\bibnamefont {Ma}},\ }\href {\doibase 10.1039/D3MH00868A} {\bibfield
  {journal} {\bibinfo  {journal} {Mater. Horiz.}\ }\textbf {\bibinfo {volume}
  {10}},\ \bibinfo {pages} {5071} (\bibinfo {year} {2023})}\BibitemShut
  {NoStop}%
\bibitem [{\citenamefont {Moriya}(1960{\natexlab{a}})}]{DM3}%
  \BibitemOpen
  \bibfield  {author} {\bibinfo {author} {\bibfnamefont {T.}~\bibnamefont
  {Moriya}},\ }\href {\doibase 10.1103/PhysRevLett.4.228} {\bibfield  {journal}
  {\bibinfo  {journal} {Phys. Rev. Lett.}\ }\textbf {\bibinfo {volume} {4}},\
  \bibinfo {pages} {228} (\bibinfo {year} {1960}{\natexlab{a}})}\BibitemShut
  {NoStop}%
\bibitem [{\citenamefont {Dzyaloshinsky}(1958)}]{DMI1}%
  \BibitemOpen
  \bibfield  {author} {\bibinfo {author} {\bibfnamefont {I.}~\bibnamefont
  {Dzyaloshinsky}},\ }\href
  {https://www.sciencedirect.com/science/article/abs/pii/0022369758900763}
  {\bibfield  {journal} {\bibinfo  {journal} {J. Phys. Chem. Solids}\ }\textbf
  {\bibinfo {volume} {4}},\ \bibinfo {pages} {241} (\bibinfo {year}
  {1958})}\BibitemShut {NoStop}%
\bibitem [{\citenamefont {Moriya}(1960{\natexlab{b}})}]{DMI2}%
  \BibitemOpen
  \bibfield  {author} {\bibinfo {author} {\bibfnamefont {T.}~\bibnamefont
  {Moriya}},\ }\href {\doibase 10.1103/PhysRev.120.91} {\bibfield  {journal}
  {\bibinfo  {journal} {Phys. Rev.}\ }\textbf {\bibinfo {volume} {120}},\
  \bibinfo {pages} {91} (\bibinfo {year} {1960}{\natexlab{b}})}\BibitemShut
  {NoStop}%
\bibitem [{\citenamefont {Huang}\ \emph {et~al.}(2017)\citenamefont {Huang},
  \citenamefont {Clark}, \citenamefont {Navarro-Moratalla}, \citenamefont
  {Klein}, \citenamefont {Cheng}, \citenamefont {Seyler}, \citenamefont
  {Zhong}, \citenamefont {Schmidgall}, \citenamefont {McGuire}, \citenamefont
  {Cobden} \emph {et~al.}}]{2D1}%
  \BibitemOpen
  \bibfield  {author} {\bibinfo {author} {\bibfnamefont {B.}~\bibnamefont
  {Huang}}, \bibinfo {author} {\bibfnamefont {G.}~\bibnamefont {Clark}},
  \bibinfo {author} {\bibfnamefont {E.}~\bibnamefont {Navarro-Moratalla}},
  \bibinfo {author} {\bibfnamefont {D.~R.}\ \bibnamefont {Klein}}, \bibinfo
  {author} {\bibfnamefont {R.}~\bibnamefont {Cheng}}, \bibinfo {author}
  {\bibfnamefont {K.~L.}\ \bibnamefont {Seyler}}, \bibinfo {author}
  {\bibfnamefont {D.}~\bibnamefont {Zhong}}, \bibinfo {author} {\bibfnamefont
  {E.}~\bibnamefont {Schmidgall}}, \bibinfo {author} {\bibfnamefont {M.~A.}\
  \bibnamefont {McGuire}}, \bibinfo {author} {\bibfnamefont {D.~H.}\
  \bibnamefont {Cobden}},  \emph {et~al.},\ }\href
  {https://www.nature.com/articles/nature22391} {\bibfield  {journal} {\bibinfo
   {journal} {Nature}\ }\textbf {\bibinfo {volume} {546}},\ \bibinfo {pages}
  {270} (\bibinfo {year} {2017})}\BibitemShut {NoStop}%
\bibitem [{\citenamefont {Gong}\ \emph {et~al.}(2017)\citenamefont {Gong},
  \citenamefont {Li}, \citenamefont {Li}, \citenamefont {Ji}, \citenamefont
  {Stern}, \citenamefont {Xia}, \citenamefont {Cao}, \citenamefont {Bao},
  \citenamefont {Wang}, \citenamefont {Wang} \emph {et~al.}}]{2D2}%
  \BibitemOpen
  \bibfield  {author} {\bibinfo {author} {\bibfnamefont {C.}~\bibnamefont
  {Gong}}, \bibinfo {author} {\bibfnamefont {L.}~\bibnamefont {Li}}, \bibinfo
  {author} {\bibfnamefont {Z.}~\bibnamefont {Li}}, \bibinfo {author}
  {\bibfnamefont {H.}~\bibnamefont {Ji}}, \bibinfo {author} {\bibfnamefont
  {A.}~\bibnamefont {Stern}}, \bibinfo {author} {\bibfnamefont
  {Y.}~\bibnamefont {Xia}}, \bibinfo {author} {\bibfnamefont {T.}~\bibnamefont
  {Cao}}, \bibinfo {author} {\bibfnamefont {W.}~\bibnamefont {Bao}}, \bibinfo
  {author} {\bibfnamefont {C.}~\bibnamefont {Wang}}, \bibinfo {author}
  {\bibfnamefont {Y.}~\bibnamefont {Wang}},  \emph {et~al.},\ }\href
  {https://www.nature.com/articles/nature22060} {\bibfield  {journal} {\bibinfo
   {journal} {Nature}\ }\textbf {\bibinfo {volume} {546}},\ \bibinfo {pages}
  {265} (\bibinfo {year} {2017})}\BibitemShut {NoStop}%
\bibitem [{\citenamefont {Deng}\ \emph {et~al.}(2018)\citenamefont {Deng},
  \citenamefont {Yu}, \citenamefont {Song}, \citenamefont {Zhang},
  \citenamefont {Wang}, \citenamefont {Sun}, \citenamefont {Yi}, \citenamefont
  {Wu}, \citenamefont {Wu}, \citenamefont {Zhu} \emph {et~al.}}]{2D3}%
  \BibitemOpen
  \bibfield  {author} {\bibinfo {author} {\bibfnamefont {Y.}~\bibnamefont
  {Deng}}, \bibinfo {author} {\bibfnamefont {Y.}~\bibnamefont {Yu}}, \bibinfo
  {author} {\bibfnamefont {Y.}~\bibnamefont {Song}}, \bibinfo {author}
  {\bibfnamefont {J.}~\bibnamefont {Zhang}}, \bibinfo {author} {\bibfnamefont
  {N.~Z.}\ \bibnamefont {Wang}}, \bibinfo {author} {\bibfnamefont
  {Z.}~\bibnamefont {Sun}}, \bibinfo {author} {\bibfnamefont {Y.}~\bibnamefont
  {Yi}}, \bibinfo {author} {\bibfnamefont {Y.~Z.}\ \bibnamefont {Wu}}, \bibinfo
  {author} {\bibfnamefont {S.}~\bibnamefont {Wu}}, \bibinfo {author}
  {\bibfnamefont {J.}~\bibnamefont {Zhu}},  \emph {et~al.},\ }\href
  {https://www.nature.com/articles/s41586-018-0626-9} {\bibfield  {journal}
  {\bibinfo  {journal} {Nature}\ }\textbf {\bibinfo {volume} {563}},\ \bibinfo
  {pages} {94} (\bibinfo {year} {2018})}\BibitemShut {NoStop}%
\bibitem [{\citenamefont {Du}\ \emph {et~al.}(2021)\citenamefont {Du},
  \citenamefont {Hasan}, \citenamefont {Castellanos-Gomez}, \citenamefont
  {Liu}, \citenamefont {Yao}, \citenamefont {Lau},\ and\ \citenamefont
  {Sun}}]{2D4}%
  \BibitemOpen
  \bibfield  {author} {\bibinfo {author} {\bibfnamefont {L.}~\bibnamefont
  {Du}}, \bibinfo {author} {\bibfnamefont {T.}~\bibnamefont {Hasan}}, \bibinfo
  {author} {\bibfnamefont {A.}~\bibnamefont {Castellanos-Gomez}}, \bibinfo
  {author} {\bibfnamefont {G.-B.}\ \bibnamefont {Liu}}, \bibinfo {author}
  {\bibfnamefont {Y.}~\bibnamefont {Yao}}, \bibinfo {author} {\bibfnamefont
  {C.~N.}\ \bibnamefont {Lau}}, \ and\ \bibinfo {author} {\bibfnamefont
  {Z.}~\bibnamefont {Sun}},\ }\href
  {https://www.nature.com/articles/s42254-020-00276-0} {\bibfield  {journal}
  {\bibinfo  {journal} {Nat. Rev. Phys.}\ }\textbf {\bibinfo {volume} {3}},\
  \bibinfo {pages} {193} (\bibinfo {year} {2021})}\BibitemShut {NoStop}%
\bibitem [{\citenamefont {Xu}\ \emph {et~al.}(2020{\natexlab{a}})\citenamefont
  {Xu}, \citenamefont {Feng}, \citenamefont {Prokhorenko}, \citenamefont
  {Nahas}, \citenamefont {Xiang},\ and\ \citenamefont {Bellaiche}}]{Janus1}%
  \BibitemOpen
  \bibfield  {author} {\bibinfo {author} {\bibfnamefont {C.}~\bibnamefont
  {Xu}}, \bibinfo {author} {\bibfnamefont {J.}~\bibnamefont {Feng}}, \bibinfo
  {author} {\bibfnamefont {S.}~\bibnamefont {Prokhorenko}}, \bibinfo {author}
  {\bibfnamefont {Y.}~\bibnamefont {Nahas}}, \bibinfo {author} {\bibfnamefont
  {H.}~\bibnamefont {Xiang}}, \ and\ \bibinfo {author} {\bibfnamefont
  {L.}~\bibnamefont {Bellaiche}},\ }\href
  {https://link.aps.org/doi/10.1103/PhysRevB.101.060404} {\bibfield  {journal}
  {\bibinfo  {journal} {Phys. Rev. B}\ }\textbf {\bibinfo {volume} {101}},\
  \bibinfo {pages} {060404(R)} (\bibinfo {year} {2020}{\natexlab{a}})}\BibitemShut
  {NoStop}%
\bibitem [{\citenamefont {Zhang}\ \emph {et~al.}(2020)\citenamefont {Zhang},
  \citenamefont {Xu}, \citenamefont {Chen}, \citenamefont {Nahas},
  \citenamefont {Prokhorenko},\ and\ \citenamefont {Bellaiche}}]{Janus2}%
  \BibitemOpen
  \bibfield  {author} {\bibinfo {author} {\bibfnamefont {Y.}~\bibnamefont
  {Zhang}}, \bibinfo {author} {\bibfnamefont {C.}~\bibnamefont {Xu}}, \bibinfo
  {author} {\bibfnamefont {P.}~\bibnamefont {Chen}}, \bibinfo {author}
  {\bibfnamefont {Y.}~\bibnamefont {Nahas}}, \bibinfo {author} {\bibfnamefont
  {S.}~\bibnamefont {Prokhorenko}}, \ and\ \bibinfo {author} {\bibfnamefont
  {L.}~\bibnamefont {Bellaiche}},\ }\href
  {https://link.aps.org/doi/10.1103/PhysRevB.102.241107} {\bibfield  {journal}
  {\bibinfo  {journal} {Phys. Rev. B}\ }\textbf {\bibinfo {volume} {102}},\
  \bibinfo {pages} {241107(R)} (\bibinfo {year} {2020})}\BibitemShut {NoStop}%
\bibitem [{\citenamefont {Hou}\ \emph {et~al.}(2022)\citenamefont {Hou},
  \citenamefont {Xue}, \citenamefont {Qiu}, \citenamefont {Wang},\ and\
  \citenamefont {Wu}}]{Janus3}%
  \BibitemOpen
  \bibfield  {author} {\bibinfo {author} {\bibfnamefont {Y.}~\bibnamefont
  {Hou}}, \bibinfo {author} {\bibfnamefont {F.}~\bibnamefont {Xue}}, \bibinfo
  {author} {\bibfnamefont {L.}~\bibnamefont {Qiu}}, \bibinfo {author}
  {\bibfnamefont {Z.}~\bibnamefont {Wang}}, \ and\ \bibinfo {author}
  {\bibfnamefont {R.}~\bibnamefont {Wu}},\ }\href
  {https://www.nature.com/articles/s41524-022-00802-x} {\bibfield  {journal}
  {\bibinfo  {journal} {npj Comput. Mater.}\ }\textbf {\bibinfo {volume} {8}},\
  \bibinfo {pages} {1} (\bibinfo {year} {2022})}\BibitemShut {NoStop}%
\bibitem [{\citenamefont {Huang}\ \emph {et~al.}(2021)\citenamefont {Huang},
  \citenamefont {Guan}, \citenamefont {Li}, \citenamefont {Wu}, \citenamefont
  {Jena},\ and\ \citenamefont {Kan}}]{E1}%
  \BibitemOpen
  \bibfield  {author} {\bibinfo {author} {\bibfnamefont {C.}~\bibnamefont
  {Huang}}, \bibinfo {author} {\bibfnamefont {J.}~\bibnamefont {Guan}},
  \bibinfo {author} {\bibfnamefont {Q.}~\bibnamefont {Li}}, \bibinfo {author}
  {\bibfnamefont {F.}~\bibnamefont {Wu}}, \bibinfo {author} {\bibfnamefont
  {P.}~\bibnamefont {Jena}}, \ and\ \bibinfo {author} {\bibfnamefont
  {E.}~\bibnamefont {Kan}},\ }\href
  {https://link.aps.org/doi/10.1103/PhysRevB.103.L140410} {\bibfield  {journal}
  {\bibinfo  {journal} {Phys. Rev. B}\ }\textbf {\bibinfo {volume} {103}},\
  \bibinfo {pages} {L140410} (\bibinfo {year} {2021})}\BibitemShut {NoStop}%
\bibitem [{\citenamefont {Liang}\ \emph {et~al.}(2020)\citenamefont {Liang},
  \citenamefont {Cui},\ and\ \citenamefont {Yang}}]{E2}%
  \BibitemOpen
  \bibfield  {author} {\bibinfo {author} {\bibfnamefont {J.}~\bibnamefont
  {Liang}}, \bibinfo {author} {\bibfnamefont {Q.}~\bibnamefont {Cui}}, \ and\
  \bibinfo {author} {\bibfnamefont {H.}~\bibnamefont {Yang}},\ }\href
  {https://link.aps.org/doi/10.1103/PhysRevB.102.220409} {\bibfield  {journal}
  {\bibinfo  {journal} {Phys. Rev. B}\ }\textbf {\bibinfo {volume} {102}},\
  \bibinfo {pages} {220409(R)} (\bibinfo {year} {2020})}\BibitemShut {NoStop}%
\bibitem [{\citenamefont {Xu}\ \emph {et~al.}(2020{\natexlab{b}})\citenamefont
  {Xu}, \citenamefont {Chen}, \citenamefont {Tan}, \citenamefont {Yang},
  \citenamefont {Xiang},\ and\ \citenamefont {Bellaiche}}]{E3}%
  \BibitemOpen
  \bibfield  {author} {\bibinfo {author} {\bibfnamefont {C.}~\bibnamefont
  {Xu}}, \bibinfo {author} {\bibfnamefont {P.}~\bibnamefont {Chen}}, \bibinfo
  {author} {\bibfnamefont {H.}~\bibnamefont {Tan}}, \bibinfo {author}
  {\bibfnamefont {Y.}~\bibnamefont {Yang}}, \bibinfo {author} {\bibfnamefont
  {H.}~\bibnamefont {Xiang}}, \ and\ \bibinfo {author} {\bibfnamefont
  {L.}~\bibnamefont {Bellaiche}},\ }\href
  {https://link.aps.org/doi/10.1103/PhysRevLett.125.037203} {\bibfield
  {journal} {\bibinfo  {journal} {Phys. Rev. Lett.}\ }\textbf {\bibinfo
  {volume} {125}},\ \bibinfo {pages} {037203} (\bibinfo {year}
  {2020}{\natexlab{b}})}\BibitemShut {NoStop}%
\bibitem [{\citenamefont {Tong}\ \emph {et~al.}(2018)\citenamefont {Tong},
  \citenamefont {Liu}, \citenamefont {Xiao},\ and\ \citenamefont
  {Yao}}]{Moire1}%
  \BibitemOpen
  \bibfield  {author} {\bibinfo {author} {\bibfnamefont {Q.}~\bibnamefont
  {Tong}}, \bibinfo {author} {\bibfnamefont {F.}~\bibnamefont {Liu}}, \bibinfo
  {author} {\bibfnamefont {J.}~\bibnamefont {Xiao}}, \ and\ \bibinfo {author}
  {\bibfnamefont {W.}~\bibnamefont {Yao}},\ }\href
  {https://pubs.acs.org/doi/10.1021/acs.nanolett.8b03315} {\bibfield  {journal}
  {\bibinfo  {journal} {Nano Lett.}\ }\textbf {\bibinfo {volume} {18}},\
  \bibinfo {pages} {7194} (\bibinfo {year} {2018})}\BibitemShut {NoStop}%
\bibitem [{\citenamefont {Hejazi}\ \emph {et~al.}(2021)\citenamefont {Hejazi},
  \citenamefont {Luo},\ and\ \citenamefont {Balents}}]{Moire2}%
  \BibitemOpen
  \bibfield  {author} {\bibinfo {author} {\bibfnamefont {K.}~\bibnamefont
  {Hejazi}}, \bibinfo {author} {\bibfnamefont {Z.-X.}\ \bibnamefont {Luo}}, \
  and\ \bibinfo {author} {\bibfnamefont {L.}~\bibnamefont {Balents}},\ }\href
  {\doibase 10.1103/PhysRevB.104.L100406} {\bibfield  {journal} {\bibinfo
  {journal} {Phys. Rev. B}\ }\textbf {\bibinfo {volume} {104}},\ \bibinfo
  {pages} {L100406} (\bibinfo {year} {2021})}\BibitemShut {NoStop}%
\bibitem [{\citenamefont {Chatterjee}\ \emph {et~al.}(2020)\citenamefont
  {Chatterjee}, \citenamefont {Bultinck},\ and\ \citenamefont
  {Zaletel}}]{Moire3}%
  \BibitemOpen
  \bibfield  {author} {\bibinfo {author} {\bibfnamefont {S.}~\bibnamefont
  {Chatterjee}}, \bibinfo {author} {\bibfnamefont {N.}~\bibnamefont
  {Bultinck}}, \ and\ \bibinfo {author} {\bibfnamefont {M.~P.}\ \bibnamefont
  {Zaletel}},\ }\href {\doibase 10.1103/PhysRevB.101.165141} {\bibfield
  {journal} {\bibinfo  {journal} {Phys. Rev. B}\ }\textbf {\bibinfo {volume}
  {101}},\ \bibinfo {pages} {165141} (\bibinfo {year} {2020})}\BibitemShut
  {NoStop}%
\bibitem [{\citenamefont {Cui}\ \emph {et~al.}(2021)\citenamefont {Cui},
  \citenamefont {Zhu}, \citenamefont {Jiang}, \citenamefont {Liang},
  \citenamefont {Yu}, \citenamefont {Cui},\ and\ \citenamefont
  {Yang}}]{Hetero1}%
  \BibitemOpen
  \bibfield  {author} {\bibinfo {author} {\bibfnamefont {Q.}~\bibnamefont
  {Cui}}, \bibinfo {author} {\bibfnamefont {Y.}~\bibnamefont {Zhu}}, \bibinfo
  {author} {\bibfnamefont {J.}~\bibnamefont {Jiang}}, \bibinfo {author}
  {\bibfnamefont {J.}~\bibnamefont {Liang}}, \bibinfo {author} {\bibfnamefont
  {D.}~\bibnamefont {Yu}}, \bibinfo {author} {\bibfnamefont {P.}~\bibnamefont
  {Cui}}, \ and\ \bibinfo {author} {\bibfnamefont {H.}~\bibnamefont {Yang}},\
  }\href {\doibase 10.1103/PhysRevResearch.3.043011} {\bibfield  {journal}
  {\bibinfo  {journal} {Phys. Rev. Res.}\ }\textbf {\bibinfo {volume} {3}},\
  \bibinfo {pages} {043011} (\bibinfo {year} {2021})}\BibitemShut {NoStop}%
\bibitem [{\citenamefont {Dou}\ \emph {et~al.}(2022)\citenamefont {Dou},
  \citenamefont {Du}, \citenamefont {Dai}, \citenamefont {Huang},\ and\
  \citenamefont {Ma}}]{Hetero2}%
  \BibitemOpen
  \bibfield  {author} {\bibinfo {author} {\bibfnamefont {K.}~\bibnamefont
  {Dou}}, \bibinfo {author} {\bibfnamefont {W.}~\bibnamefont {Du}}, \bibinfo
  {author} {\bibfnamefont {Y.}~\bibnamefont {Dai}}, \bibinfo {author}
  {\bibfnamefont {B.}~\bibnamefont {Huang}}, \ and\ \bibinfo {author}
  {\bibfnamefont {Y.}~\bibnamefont {Ma}},\ }\href
  {https://link.aps.org/doi/10.1103/PhysRevB.105.205427} {\bibfield  {journal}
  {\bibinfo  {journal} {Phys. Rev. B}\ }\textbf {\bibinfo {volume} {105}},\
  \bibinfo {pages} {205427} (\bibinfo {year} {2022})}\BibitemShut {NoStop}%
\bibitem [{\citenamefont {Wu}\ \emph {et~al.}(2022{\natexlab{a}})\citenamefont
  {Wu}, \citenamefont {Francisco}, \citenamefont {Chen}, \citenamefont {Wang},
  \citenamefont {Zhang}, \citenamefont {Wan}, \citenamefont {Han},
  \citenamefont {Chi}, \citenamefont {Hou}, \citenamefont {Lodesani} \emph
  {et~al.}}]{Hetero3}%
  \BibitemOpen
  \bibfield  {author} {\bibinfo {author} {\bibfnamefont {Y.}~\bibnamefont
  {Wu}}, \bibinfo {author} {\bibfnamefont {B.}~\bibnamefont {Francisco}},
  \bibinfo {author} {\bibfnamefont {Z.}~\bibnamefont {Chen}}, \bibinfo {author}
  {\bibfnamefont {W.}~\bibnamefont {Wang}}, \bibinfo {author} {\bibfnamefont
  {Y.}~\bibnamefont {Zhang}}, \bibinfo {author} {\bibfnamefont
  {C.}~\bibnamefont {Wan}}, \bibinfo {author} {\bibfnamefont {X.}~\bibnamefont
  {Han}}, \bibinfo {author} {\bibfnamefont {H.}~\bibnamefont {Chi}}, \bibinfo
  {author} {\bibfnamefont {Y.}~\bibnamefont {Hou}}, \bibinfo {author}
  {\bibfnamefont {A.}~\bibnamefont {Lodesani}},  \emph {et~al.},\ }\href
  {https://onlinelibrary.wiley.com/doi/full/10.1002/adma.202110583} {\bibfield
  {journal} {\bibinfo  {journal} {Adv. Mater.}\ }\textbf {\bibinfo {volume}
  {34}},\ \bibinfo {pages} {2110583} (\bibinfo {year}
  {2022}{\natexlab{a}})}\BibitemShut {NoStop}%
\bibitem [{\citenamefont {Abuawwad}\ \emph {et~al.}(2022)\citenamefont
  {Abuawwad}, \citenamefont {dos Santos~Dias}, \citenamefont {Abusara},\ and\
  \citenamefont {Lounis}}]{fru1}%
  \BibitemOpen
  \bibfield  {author} {\bibinfo {author} {\bibfnamefont {N.}~\bibnamefont
  {Abuawwad}}, \bibinfo {author} {\bibfnamefont {M.}~\bibnamefont {dos
  Santos~Dias}}, \bibinfo {author} {\bibfnamefont {H.}~\bibnamefont {Abusara}},
  \ and\ \bibinfo {author} {\bibfnamefont {S.}~\bibnamefont {Lounis}},\ }\href
  {\doibase 10.1103/PhysRevLett.115.267210} {\bibfield  {journal} {\bibinfo
  {journal} {J. Phys.: Condens. Matter}\ }\textbf {\bibinfo {volume} {34}},\
  \bibinfo {pages} {454001} (\bibinfo {year} {2022})}\BibitemShut {NoStop}%
\bibitem [{\citenamefont {Hu}\ \emph {et~al.}(2017)\citenamefont {Hu},
  \citenamefont {Chi}, \citenamefont {Li}, \citenamefont {Liu},\ and\
  \citenamefont {Du}}]{fru2}%
  \BibitemOpen
  \bibfield  {author} {\bibinfo {author} {\bibfnamefont {Y.}~\bibnamefont
  {Hu}}, \bibinfo {author} {\bibfnamefont {X.}~\bibnamefont {Chi}}, \bibinfo
  {author} {\bibfnamefont {X.}~\bibnamefont {Li}}, \bibinfo {author}
  {\bibfnamefont {Y.}~\bibnamefont {Liu}}, \ and\ \bibinfo {author}
  {\bibfnamefont {A.}~\bibnamefont {Du}},\ }\href {\doibase
  https://www.nature.com/articles/s41598-017-16348-8} {\bibfield  {journal}
  {\bibinfo  {journal} {Sci Rep}\ }\textbf {\bibinfo {volume} {7}},\ \bibinfo
  {pages} {16079} (\bibinfo {year} {2017})}\BibitemShut {NoStop}%
\bibitem [{\citenamefont {von Malottki}\ \emph {et~al.}(2017)\citenamefont {von
  Malottki}, \citenamefont {Dup{\'e}}, \citenamefont {Bessarab}, \citenamefont
  {Delin},\ and\ \citenamefont {Heinze}}]{fru3}%
  \BibitemOpen
  \bibfield  {author} {\bibinfo {author} {\bibfnamefont {S.}~\bibnamefont {von
  Malottki}}, \bibinfo {author} {\bibfnamefont {B.}~\bibnamefont {Dup{\'e}}},
  \bibinfo {author} {\bibfnamefont {P.~F.}\ \bibnamefont {Bessarab}}, \bibinfo
  {author} {\bibfnamefont {A.}~\bibnamefont {Delin}}, \ and\ \bibinfo {author}
  {\bibfnamefont {S.}~\bibnamefont {Heinze}},\ }\href
  {https://www.nature.com/articles/s41598-017-12525-x} {\bibfield  {journal}
  {\bibinfo  {journal} {Sci Rep}\ }\textbf {\bibinfo {volume} {7}},\ \bibinfo
  {pages} {12299} (\bibinfo {year} {2017})}\BibitemShut {NoStop}%
\bibitem [{\citenamefont {Shen}\ \emph {et~al.}(2023)\citenamefont {Shen},
  \citenamefont {Dong},\ and\ \citenamefont {Yao}}]{MAE1}%
  \BibitemOpen
  \bibfield  {author} {\bibinfo {author} {\bibfnamefont {Z.}~\bibnamefont
  {Shen}}, \bibinfo {author} {\bibfnamefont {S.}~\bibnamefont {Dong}}, \ and\
  \bibinfo {author} {\bibfnamefont {X.}~\bibnamefont {Yao}},\ }\href
  {https://link.aps.org/doi/10.1103/PhysRevB.108.L140412} {\bibfield  {journal}
  {\bibinfo  {journal} {Phys. Rev. B}\ }\textbf {\bibinfo {volume} {108}},\
  \bibinfo {pages} {L140412} (\bibinfo {year} {2023})}\BibitemShut {NoStop}%
\bibitem [{\citenamefont {Amoroso}\ \emph {et~al.}(2020)\citenamefont
  {Amoroso}, \citenamefont {Barone},\ and\ \citenamefont {Picozzi}}]{Kitaev2}%
  \BibitemOpen
  \bibfield  {author} {\bibinfo {author} {\bibfnamefont {D.}~\bibnamefont
  {Amoroso}}, \bibinfo {author} {\bibfnamefont {P.}~\bibnamefont {Barone}}, \
  and\ \bibinfo {author} {\bibfnamefont {S.}~\bibnamefont {Picozzi}},\ }\href
  {https://www.nature.com/articles/s41467-020-19535-w} {\bibfield  {journal}
  {\bibinfo  {journal} {Nat Commun}\ }\textbf {\bibinfo {volume} {11}},\
  \bibinfo {pages} {5784} (\bibinfo {year} {2020})}\BibitemShut {NoStop}%
\bibitem [{\citenamefont {Hayami}\ \emph {et~al.}(2017)\citenamefont {Hayami},
  \citenamefont {Ozawa},\ and\ \citenamefont {Motome}}]{Novel1}%
  \BibitemOpen
  \bibfield  {author} {\bibinfo {author} {\bibfnamefont {S.}~\bibnamefont
  {Hayami}}, \bibinfo {author} {\bibfnamefont {R.}~\bibnamefont {Ozawa}}, \
  and\ \bibinfo {author} {\bibfnamefont {Y.}~\bibnamefont {Motome}},\ }\href
  {\doibase 10.1103/PhysRevB.95.224424} {\bibfield  {journal} {\bibinfo
  {journal} {Phys. Rev. B}\ }\textbf {\bibinfo {volume} {95}},\ \bibinfo
  {pages} {224424} (\bibinfo {year} {2017})}\BibitemShut {NoStop}%
\bibitem [{\citenamefont {Ni}\ \emph {et~al.}(2021)\citenamefont {Ni},
  \citenamefont {Li}, \citenamefont {Amoroso}, \citenamefont {He},
  \citenamefont {Feng}, \citenamefont {Kan}, \citenamefont {Picozzi},\ and\
  \citenamefont {Xiang}}]{Novel2}%
  \BibitemOpen
  \bibfield  {author} {\bibinfo {author} {\bibfnamefont {J.~Y.}\ \bibnamefont
  {Ni}}, \bibinfo {author} {\bibfnamefont {X.~Y.}\ \bibnamefont {Li}}, \bibinfo
  {author} {\bibfnamefont {D.}~\bibnamefont {Amoroso}}, \bibinfo {author}
  {\bibfnamefont {X.}~\bibnamefont {He}}, \bibinfo {author} {\bibfnamefont
  {J.~S.}\ \bibnamefont {Feng}}, \bibinfo {author} {\bibfnamefont {E.~J.}\
  \bibnamefont {Kan}}, \bibinfo {author} {\bibfnamefont {S.}~\bibnamefont
  {Picozzi}}, \ and\ \bibinfo {author} {\bibfnamefont {H.~J.}\ \bibnamefont
  {Xiang}},\ }\href {\doibase 10.1103/PhysRevLett.127.247204} {\bibfield
  {journal} {\bibinfo  {journal} {Phys. Rev. Lett.}\ }\textbf {\bibinfo
  {volume} {127}},\ \bibinfo {pages} {247204} (\bibinfo {year}
  {2021})}\BibitemShut {NoStop}%
\bibitem [{\citenamefont {Xu}\ \emph {et~al.}(2022)\citenamefont {Xu},
  \citenamefont {Li}, \citenamefont {Chen}, \citenamefont {Zhang},
  \citenamefont {Xiang},\ and\ \citenamefont {Bellaiche}}]{Novel3}%
  \BibitemOpen
  \bibfield  {author} {\bibinfo {author} {\bibfnamefont {C.}~\bibnamefont
  {Xu}}, \bibinfo {author} {\bibfnamefont {X.}~\bibnamefont {Li}}, \bibinfo
  {author} {\bibfnamefont {P.}~\bibnamefont {Chen}}, \bibinfo {author}
  {\bibfnamefont {Y.}~\bibnamefont {Zhang}}, \bibinfo {author} {\bibfnamefont
  {H.}~\bibnamefont {Xiang}}, \ and\ \bibinfo {author} {\bibfnamefont
  {L.}~\bibnamefont {Bellaiche}},\ }\href
  {https://onlinelibrary.wiley.com/doi/full/10.1002/adma.202107779} {\bibfield
  {journal} {\bibinfo  {journal} {Adv. Mater.}\ }\textbf {\bibinfo {volume}
  {34}},\ \bibinfo {pages} {2107779} (\bibinfo {year} {2022})}\BibitemShut
  {NoStop}%
\bibitem [{\citenamefont {Haldar}\ \emph {et~al.}(2021)\citenamefont {Haldar},
  \citenamefont {Meyer}, \citenamefont {Kubetzka},\ and\ \citenamefont
  {Heinze}}]{Novel4}%
  \BibitemOpen
  \bibfield  {author} {\bibinfo {author} {\bibfnamefont {S.}~\bibnamefont
  {Haldar}}, \bibinfo {author} {\bibfnamefont {S.}~\bibnamefont {Meyer}},
  \bibinfo {author} {\bibfnamefont {A.}~\bibnamefont {Kubetzka}}, \ and\
  \bibinfo {author} {\bibfnamefont {S.}~\bibnamefont {Heinze}},\ }\href
  {\doibase 10.1103/PhysRevB.104.L180404} {\bibfield  {journal} {\bibinfo
  {journal} {Phys. Rev. B}\ }\textbf {\bibinfo {volume} {104}},\ \bibinfo
  {pages} {L180404} (\bibinfo {year} {2021})}\BibitemShut {NoStop}%
\bibitem [{\citenamefont {Gutzeit}\ \emph {et~al.}(2021)\citenamefont
  {Gutzeit}, \citenamefont {Haldar}, \citenamefont {Meyer},\ and\ \citenamefont
  {Heinze}}]{Novel5}%
  \BibitemOpen
  \bibfield  {author} {\bibinfo {author} {\bibfnamefont {M.}~\bibnamefont
  {Gutzeit}}, \bibinfo {author} {\bibfnamefont {S.}~\bibnamefont {Haldar}},
  \bibinfo {author} {\bibfnamefont {S.}~\bibnamefont {Meyer}}, \ and\ \bibinfo
  {author} {\bibfnamefont {S.}~\bibnamefont {Heinze}},\ }\href
  {https://link.aps.org/doi/10.1103/PhysRevB.104.024420} {\bibfield  {journal}
  {\bibinfo  {journal} {Phys. Rev. B}\ }\textbf {\bibinfo {volume} {104}},\
  \bibinfo {pages} {024420} (\bibinfo {year} {2021})}\BibitemShut {NoStop}%
\bibitem [{\citenamefont {Hoffmann}\ and\ \citenamefont
  {Bl\"ugel}(2020)}]{Novel6}%
  \BibitemOpen
  \bibfield  {author} {\bibinfo {author} {\bibfnamefont {M.}~\bibnamefont
  {Hoffmann}}\ and\ \bibinfo {author} {\bibfnamefont {S.}~\bibnamefont
  {Bl\"ugel}},\ }\href {https://link.aps.org/doi/10.1103/PhysRevB.101.024418}
  {\bibfield  {journal} {\bibinfo  {journal} {Phys. Rev. B}\ }\textbf {\bibinfo
  {volume} {101}},\ \bibinfo {pages} {024418} (\bibinfo {year}
  {2020})}\BibitemShut {NoStop}%
\bibitem [{\citenamefont {Mankovsky}\ \emph {et~al.}(2021)\citenamefont
  {Mankovsky}, \citenamefont {Polesya},\ and\ \citenamefont {Ebert}}]{Novel7}%
  \BibitemOpen
  \bibfield  {author} {\bibinfo {author} {\bibfnamefont {S.}~\bibnamefont
  {Mankovsky}}, \bibinfo {author} {\bibfnamefont {S.}~\bibnamefont {Polesya}},
  \ and\ \bibinfo {author} {\bibfnamefont {H.}~\bibnamefont {Ebert}},\ }\href
  {https://link.aps.org/doi/10.1103/PhysRevB.104.054418} {\bibfield  {journal}
  {\bibinfo  {journal} {Phys. Rev. B}\ }\textbf {\bibinfo {volume} {104}},\
  \bibinfo {pages} {054418} (\bibinfo {year} {2021})}\BibitemShut {NoStop}%
\bibitem [{\citenamefont {Brinker}\ \emph {et~al.}(2020)\citenamefont
  {Brinker}, \citenamefont {dos Santos~Dias},\ and\ \citenamefont
  {Lounis}}]{Novel8}%
  \BibitemOpen
  \bibfield  {author} {\bibinfo {author} {\bibfnamefont {S.}~\bibnamefont
  {Brinker}}, \bibinfo {author} {\bibfnamefont {M.}~\bibnamefont {dos
  Santos~Dias}}, \ and\ \bibinfo {author} {\bibfnamefont {S.}~\bibnamefont
  {Lounis}},\ }\href
  {https://link.aps.org/doi/10.1103/PhysRevResearch.2.033240} {\bibfield
  {journal} {\bibinfo  {journal} {Phys. Rev. Res.}\ }\textbf {\bibinfo {volume}
  {2}},\ \bibinfo {pages} {033240} (\bibinfo {year} {2020})}\BibitemShut
  {NoStop}%
\bibitem [{\citenamefont {dos Santos~Dias}\ \emph {et~al.}(2021)\citenamefont
  {dos Santos~Dias}, \citenamefont {Brinker}, \citenamefont {L\'aszl\'offy},
  \citenamefont {Ny\'ari}, \citenamefont {Bl\"ugel}, \citenamefont {Szunyogh},\
  and\ \citenamefont {Lounis}}]{Novel9}%
  \BibitemOpen
  \bibfield  {author} {\bibinfo {author} {\bibfnamefont {M.}~\bibnamefont {dos
  Santos~Dias}}, \bibinfo {author} {\bibfnamefont {S.}~\bibnamefont {Brinker}},
  \bibinfo {author} {\bibfnamefont {A.}~\bibnamefont {L\'aszl\'offy}}, \bibinfo
  {author} {\bibfnamefont {B.}~\bibnamefont {Ny\'ari}}, \bibinfo {author}
  {\bibfnamefont {S.}~\bibnamefont {Bl\"ugel}}, \bibinfo {author}
  {\bibfnamefont {L.}~\bibnamefont {Szunyogh}}, \ and\ \bibinfo {author}
  {\bibfnamefont {S.}~\bibnamefont {Lounis}},\ }\href
  {https://link.aps.org/doi/10.1103/PhysRevB.103.L140408} {\bibfield  {journal}
  {\bibinfo  {journal} {Phys. Rev. B}\ }\textbf {\bibinfo {volume} {103}},\
  \bibinfo {pages} {L140408} (\bibinfo {year} {2021})}\BibitemShut {NoStop}%
\bibitem [{\citenamefont {Lounis}(2020)}]{Novel10}%
  \BibitemOpen
  \bibfield  {author} {\bibinfo {author} {\bibfnamefont {S.}~\bibnamefont
  {Lounis}},\ }\href
  {https://iopscience.iop.org/article/10.1088/1367-2630/abb514} {\bibfield
  {journal} {\bibinfo  {journal} {New J. Phys.}\ }\textbf {\bibinfo {volume}
  {22}},\ \bibinfo {pages} {103003} (\bibinfo {year} {2020})}\BibitemShut
  {NoStop}%
\bibitem [{\citenamefont {Brinker}\ \emph {et~al.}(2019)\citenamefont
  {Brinker}, \citenamefont {Dias},\ and\ \citenamefont {Lounis}}]{Novel11}%
  \BibitemOpen
  \bibfield  {author} {\bibinfo {author} {\bibfnamefont {S.}~\bibnamefont
  {Brinker}}, \bibinfo {author} {\bibfnamefont {M.~d.~S.}\ \bibnamefont
  {Dias}}, \ and\ \bibinfo {author} {\bibfnamefont {S.}~\bibnamefont
  {Lounis}},\ }\href
  {https://iopscience.iop.org/article/10.1088/1367-2630/ab35c9} {\bibfield
  {journal} {\bibinfo  {journal} {New J. Phys.}\ }\textbf {\bibinfo {volume}
  {21}},\ \bibinfo {pages} {083015} (\bibinfo {year} {2019})}\BibitemShut
  {NoStop}%
\bibitem [{\citenamefont {Grytsiuk}\ \emph {et~al.}(2020)\citenamefont
  {Grytsiuk}, \citenamefont {Hanke}, \citenamefont {Hoffmann}, \citenamefont
  {Bouaziz}, \citenamefont {Gomonay}, \citenamefont {Bihlmayer}, \citenamefont
  {Lounis}, \citenamefont {Mokrousov},\ and\ \citenamefont
  {Bl{\"u}gel}}]{Novel12}%
  \BibitemOpen
  \bibfield  {author} {\bibinfo {author} {\bibfnamefont {S.}~\bibnamefont
  {Grytsiuk}}, \bibinfo {author} {\bibfnamefont {J.-P.}\ \bibnamefont {Hanke}},
  \bibinfo {author} {\bibfnamefont {M.}~\bibnamefont {Hoffmann}}, \bibinfo
  {author} {\bibfnamefont {J.}~\bibnamefont {Bouaziz}}, \bibinfo {author}
  {\bibfnamefont {O.}~\bibnamefont {Gomonay}}, \bibinfo {author} {\bibfnamefont
  {G.}~\bibnamefont {Bihlmayer}}, \bibinfo {author} {\bibfnamefont
  {S.}~\bibnamefont {Lounis}}, \bibinfo {author} {\bibfnamefont
  {Y.}~\bibnamefont {Mokrousov}}, \ and\ \bibinfo {author} {\bibfnamefont
  {S.}~\bibnamefont {Bl{\"u}gel}},\ }\href
  {https://www.nature.com/articles/s41467-019-14030-3} {\bibfield  {journal}
  {\bibinfo  {journal} {Nat. Commun.}\ }\textbf {\bibinfo {volume} {11}},\
  \bibinfo {pages} {511} (\bibinfo {year} {2020})}\BibitemShut {NoStop}%
\bibitem [{\citenamefont {Simon}\ \emph {et~al.}(2020)\citenamefont {Simon},
  \citenamefont {Donges}, \citenamefont {Szunyogh},\ and\ \citenamefont
  {Nowak}}]{Novelexp1}%
  \BibitemOpen
  \bibfield  {author} {\bibinfo {author} {\bibfnamefont {E.}~\bibnamefont
  {Simon}}, \bibinfo {author} {\bibfnamefont {A.}~\bibnamefont {Donges}},
  \bibinfo {author} {\bibfnamefont {L.}~\bibnamefont {Szunyogh}}, \ and\
  \bibinfo {author} {\bibfnamefont {U.}~\bibnamefont {Nowak}},\ }\href
  {https://link.aps.org/doi/10.1103/PhysRevMaterials.4.084408} {\bibfield
  {journal} {\bibinfo  {journal} {Phys. Rev. Materials}\ }\textbf {\bibinfo
  {volume} {4}},\ \bibinfo {pages} {084408} (\bibinfo {year}
  {2020})}\BibitemShut {NoStop}%
\bibitem [{\citenamefont {Romming}\ \emph {et~al.}(2018)\citenamefont
  {Romming}, \citenamefont {Pralow}, \citenamefont {Kubetzka}, \citenamefont
  {Hoffmann}, \citenamefont {von Malottki}, \citenamefont {Meyer},
  \citenamefont {Dup\'e}, \citenamefont {Wiesendanger}, \citenamefont {von
  Bergmann},\ and\ \citenamefont {Heinze}}]{Novelexp2}%
  \BibitemOpen
  \bibfield  {author} {\bibinfo {author} {\bibfnamefont {N.}~\bibnamefont
  {Romming}}, \bibinfo {author} {\bibfnamefont {H.}~\bibnamefont {Pralow}},
  \bibinfo {author} {\bibfnamefont {A.}~\bibnamefont {Kubetzka}}, \bibinfo
  {author} {\bibfnamefont {M.}~\bibnamefont {Hoffmann}}, \bibinfo {author}
  {\bibfnamefont {S.}~\bibnamefont {von Malottki}}, \bibinfo {author}
  {\bibfnamefont {S.}~\bibnamefont {Meyer}}, \bibinfo {author} {\bibfnamefont
  {B.}~\bibnamefont {Dup\'e}}, \bibinfo {author} {\bibfnamefont
  {R.}~\bibnamefont {Wiesendanger}}, \bibinfo {author} {\bibfnamefont
  {K.}~\bibnamefont {von Bergmann}}, \ and\ \bibinfo {author} {\bibfnamefont
  {S.}~\bibnamefont {Heinze}},\ }\href
  {https://link.aps.org/doi/10.1103/PhysRevLett.120.207201} {\bibfield
  {journal} {\bibinfo  {journal} {Phys. Rev. Lett.}\ }\textbf {\bibinfo
  {volume} {120}},\ \bibinfo {pages} {207201} (\bibinfo {year}
  {2018})}\BibitemShut {NoStop}%
\bibitem [{\citenamefont {Kr\"onlein}\ \emph {et~al.}(2018)\citenamefont
  {Kr\"onlein}, \citenamefont {Schmitt}, \citenamefont {Hoffmann},
  \citenamefont {Kemmer}, \citenamefont {Seubert}, \citenamefont {Vogt},
  \citenamefont {K\"uspert}, \citenamefont {B\"ohme}, \citenamefont {Alonazi},
  \citenamefont {K\"ugel}, \citenamefont {Albrithen}, \citenamefont {Bode},
  \citenamefont {Bihlmayer},\ and\ \citenamefont {Bl\"ugel}}]{Novelexp3}%
  \BibitemOpen
  \bibfield  {author} {\bibinfo {author} {\bibfnamefont {A.}~\bibnamefont
  {Kr\"onlein}}, \bibinfo {author} {\bibfnamefont {M.}~\bibnamefont {Schmitt}},
  \bibinfo {author} {\bibfnamefont {M.}~\bibnamefont {Hoffmann}}, \bibinfo
  {author} {\bibfnamefont {J.}~\bibnamefont {Kemmer}}, \bibinfo {author}
  {\bibfnamefont {N.}~\bibnamefont {Seubert}}, \bibinfo {author} {\bibfnamefont
  {M.}~\bibnamefont {Vogt}}, \bibinfo {author} {\bibfnamefont {J.}~\bibnamefont
  {K\"uspert}}, \bibinfo {author} {\bibfnamefont {M.}~\bibnamefont {B\"ohme}},
  \bibinfo {author} {\bibfnamefont {B.}~\bibnamefont {Alonazi}}, \bibinfo
  {author} {\bibfnamefont {J.}~\bibnamefont {K\"ugel}}, \bibinfo {author}
  {\bibfnamefont {H.~A.}\ \bibnamefont {Albrithen}}, \bibinfo {author}
  {\bibfnamefont {M.}~\bibnamefont {Bode}}, \bibinfo {author} {\bibfnamefont
  {G.}~\bibnamefont {Bihlmayer}}, \ and\ \bibinfo {author} {\bibfnamefont
  {S.}~\bibnamefont {Bl\"ugel}},\ }\href
  {https://link.aps.org/doi/10.1103/PhysRevLett.120.207202} {\bibfield
  {journal} {\bibinfo  {journal} {Phys. Rev. Lett.}\ }\textbf {\bibinfo
  {volume} {120}},\ \bibinfo {pages} {207202} (\bibinfo {year}
  {2018})}\BibitemShut {NoStop}%
\bibitem [{\citenamefont {Spethmann}\ \emph {et~al.}(2020)\citenamefont
  {Spethmann}, \citenamefont {Meyer}, \citenamefont {von Bergmann},
  \citenamefont {Wiesendanger}, \citenamefont {Heinze},\ and\ \citenamefont
  {Kubetzka}}]{Novelexp4}%
  \BibitemOpen
  \bibfield  {author} {\bibinfo {author} {\bibfnamefont {J.}~\bibnamefont
  {Spethmann}}, \bibinfo {author} {\bibfnamefont {S.}~\bibnamefont {Meyer}},
  \bibinfo {author} {\bibfnamefont {K.}~\bibnamefont {von Bergmann}}, \bibinfo
  {author} {\bibfnamefont {R.}~\bibnamefont {Wiesendanger}}, \bibinfo {author}
  {\bibfnamefont {S.}~\bibnamefont {Heinze}}, \ and\ \bibinfo {author}
  {\bibfnamefont {A.}~\bibnamefont {Kubetzka}},\ }\href
  {https://link.aps.org/doi/10.1103/PhysRevLett.124.227203} {\bibfield
  {journal} {\bibinfo  {journal} {Phys. Rev. Lett.}\ }\textbf {\bibinfo
  {volume} {124}},\ \bibinfo {pages} {227203} (\bibinfo {year}
  {2020})}\BibitemShut {NoStop}%
\bibitem [{\citenamefont {Li}\ \emph {et~al.}(2023{\natexlab{a}})\citenamefont
  {Li}, \citenamefont {Yu}, \citenamefont {Liang}, \citenamefont {Ga},\ and\
  \citenamefont {Yang}}]{HOIeff}%
  \BibitemOpen
  \bibfield  {author} {\bibinfo {author} {\bibfnamefont {P.}~\bibnamefont
  {Li}}, \bibinfo {author} {\bibfnamefont {D.}~\bibnamefont {Yu}}, \bibinfo
  {author} {\bibfnamefont {J.}~\bibnamefont {Liang}}, \bibinfo {author}
  {\bibfnamefont {Y.}~\bibnamefont {Ga}}, \ and\ \bibinfo {author}
  {\bibfnamefont {H.}~\bibnamefont {Yang}},\ }\href
  {https://link.aps.org/doi/10.1103/PhysRevB.107.054408} {\bibfield  {journal}
  {\bibinfo  {journal} {Phys. Rev. B}\ }\textbf {\bibinfo {volume} {107}},\
  \bibinfo {pages} {054408} (\bibinfo {year} {2023}{\natexlab{a}})}\BibitemShut
  {NoStop}%
\bibitem [{\citenamefont {Paul}\ \emph {et~al.}(2020)\citenamefont {Paul},
  \citenamefont {Haldar}, \citenamefont {Von~Malottki},\ and\ \citenamefont
  {Heinze}}]{HOIeff2}%
  \BibitemOpen
  \bibfield  {author} {\bibinfo {author} {\bibfnamefont {S.}~\bibnamefont
  {Paul}}, \bibinfo {author} {\bibfnamefont {S.}~\bibnamefont {Haldar}},
  \bibinfo {author} {\bibfnamefont {S.}~\bibnamefont {Von~Malottki}}, \ and\
  \bibinfo {author} {\bibfnamefont {S.}~\bibnamefont {Heinze}},\ }\href
  {https://www.nature.com/articles/s41467-020-18473-x} {\bibfield  {journal}
  {\bibinfo  {journal} {Nat Commun}\ }\textbf {\bibinfo {volume} {11}},\
  \bibinfo {pages} {4756} (\bibinfo {year} {2020})}\BibitemShut {NoStop}%
\bibitem [{\citenamefont {Paul}\ and\ \citenamefont {Heinze}(2022)}]{HOIeff3}%
  \BibitemOpen
  \bibfield  {author} {\bibinfo {author} {\bibfnamefont {S.}~\bibnamefont
  {Paul}}\ and\ \bibinfo {author} {\bibfnamefont {S.}~\bibnamefont {Heinze}},\
  }\href {https://www.nature.com/articles/s41524-022-00785-9} {\bibfield
  {journal} {\bibinfo  {journal} {npj Comput Mater}\ }\textbf {\bibinfo
  {volume} {8}},\ \bibinfo {pages} {105} (\bibinfo {year} {2022})}\BibitemShut
  {NoStop}%
\bibitem [{\citenamefont {Kobayashi}\ \emph {et~al.}(2016)\citenamefont
  {Kobayashi}, \citenamefont {Ueda}, \citenamefont {Michioka},\ and\
  \citenamefont {Yoshimura}}]{ACT4}%
  \BibitemOpen
  \bibfield  {author} {\bibinfo {author} {\bibfnamefont {S.}~\bibnamefont
  {Kobayashi}}, \bibinfo {author} {\bibfnamefont {H.}~\bibnamefont {Ueda}},
  \bibinfo {author} {\bibfnamefont {C.}~\bibnamefont {Michioka}}, \ and\
  \bibinfo {author} {\bibfnamefont {K.}~\bibnamefont {Yoshimura}},\ }\href
  {https://pubs.acs.org/doi/10.1021/acsaelm.0c00686} {\bibfield  {journal}
  {\bibinfo  {journal} {Inorg. Chem.}\ }\textbf {\bibinfo {volume} {55}},\
  \bibinfo {pages} {7407} (\bibinfo {year} {2016})}\BibitemShut {NoStop}%
\bibitem [{\citenamefont {Nocerino}\ \emph {et~al.}(2022)\citenamefont
  {Nocerino}, \citenamefont {Witteveen}, \citenamefont {Kobayashi},
  \citenamefont {Forslund}, \citenamefont {Matsubara}, \citenamefont {Zubayer},
  \citenamefont {Mazza}, \citenamefont {Kawaguchi}, \citenamefont {Hoshikawa},
  \citenamefont {Umegaki} \emph {et~al.}}]{ACT6}%
  \BibitemOpen
  \bibfield  {author} {\bibinfo {author} {\bibfnamefont {E.}~\bibnamefont
  {Nocerino}}, \bibinfo {author} {\bibfnamefont {C.}~\bibnamefont {Witteveen}},
  \bibinfo {author} {\bibfnamefont {S.}~\bibnamefont {Kobayashi}}, \bibinfo
  {author} {\bibfnamefont {O.~K.}\ \bibnamefont {Forslund}}, \bibinfo {author}
  {\bibfnamefont {N.}~\bibnamefont {Matsubara}}, \bibinfo {author}
  {\bibfnamefont {A.}~\bibnamefont {Zubayer}}, \bibinfo {author} {\bibfnamefont
  {F.}~\bibnamefont {Mazza}}, \bibinfo {author} {\bibfnamefont
  {S.}~\bibnamefont {Kawaguchi}}, \bibinfo {author} {\bibfnamefont
  {A.}~\bibnamefont {Hoshikawa}}, \bibinfo {author} {\bibfnamefont
  {I.}~\bibnamefont {Umegaki}},  \emph {et~al.},\ }\href
  {https://www.nature.com/articles/s41598-022-25921-9} {\bibfield  {journal}
  {\bibinfo  {journal} {Sci Rep}\ }\textbf {\bibinfo {volume} {12}},\ \bibinfo
  {pages} {21657} (\bibinfo {year} {2022})}\BibitemShut {NoStop}%
\bibitem [{\citenamefont {Xu}\ \emph {et~al.}(2020{\natexlab{c}})\citenamefont
  {Xu}, \citenamefont {Ali}, \citenamefont {Jin}, \citenamefont {Wu},\ and\
  \citenamefont {Xu}}]{ACT3}%
  \BibitemOpen
  \bibfield  {author} {\bibinfo {author} {\bibfnamefont {W.}~\bibnamefont
  {Xu}}, \bibinfo {author} {\bibfnamefont {S.}~\bibnamefont {Ali}}, \bibinfo
  {author} {\bibfnamefont {Y.}~\bibnamefont {Jin}}, \bibinfo {author}
  {\bibfnamefont {X.}~\bibnamefont {Wu}}, \ and\ \bibinfo {author}
  {\bibfnamefont {H.}~\bibnamefont {Xu}},\ }\href
  {https://pubs.acs.org/doi/10.1021/acsaelm.0c00686} {\bibfield  {journal}
  {\bibinfo  {journal} {ACS Appl. Electron. Mater.}\ }\textbf {\bibinfo
  {volume} {2}},\ \bibinfo {pages} {3853} (\bibinfo {year}
  {2020}{\natexlab{c}})}\BibitemShut {NoStop}%
\bibitem [{\citenamefont {Li}\ \emph {et~al.}(2022{\natexlab{b}})\citenamefont
  {Li}, \citenamefont {Cui}, \citenamefont {Ga}, \citenamefont {Liang},\ and\
  \citenamefont {Yang}}]{ACT5}%
  \BibitemOpen
  \bibfield  {author} {\bibinfo {author} {\bibfnamefont {P.}~\bibnamefont
  {Li}}, \bibinfo {author} {\bibfnamefont {Q.}~\bibnamefont {Cui}}, \bibinfo
  {author} {\bibfnamefont {Y.}~\bibnamefont {Ga}}, \bibinfo {author}
  {\bibfnamefont {J.}~\bibnamefont {Liang}}, \ and\ \bibinfo {author}
  {\bibfnamefont {H.}~\bibnamefont {Yang}},\ }\href
  {https://link.aps.org/doi/10.1103/PhysRevB.106.024419} {\bibfield  {journal}
  {\bibinfo  {journal} {Phys. Rev. B}\ }\textbf {\bibinfo {volume} {106}},\
  \bibinfo {pages} {024419} (\bibinfo {year} {2022}{\natexlab{b}})}\BibitemShut
  {NoStop}%
\bibitem [{SM()}]{SM}%
  \BibitemOpen
  \href@noop {} {}\bibinfo {note} {See Supplementary Materials for more details
  about the absorption site of Li atoms, test of spin models, and more details
  about parameters of spin models, which includes Refs.\cite{ACT3,ACT5}.}\BibitemShut
  {Stop}%
\bibitem [{\citenamefont {Perdew}\ \emph {et~al.}(1996)\citenamefont {Perdew},
  \citenamefont {Burke},\ and\ \citenamefont {Ernzerhof}}]{PBE}%
  \BibitemOpen
  \bibfield  {author} {\bibinfo {author} {\bibfnamefont {J.~P.}\ \bibnamefont
  {Perdew}}, \bibinfo {author} {\bibfnamefont {K.}~\bibnamefont {Burke}}, \
  and\ \bibinfo {author} {\bibfnamefont {M.}~\bibnamefont {Ernzerhof}},\ }\href
  {https://link.aps.org/doi/10.1103/PhysRevLett.77.3865} {\bibfield  {journal}
  {\bibinfo  {journal} {Phys. Rev. Lett.}\ }\textbf {\bibinfo {volume} {77}},\
  \bibinfo {pages} {3865} (\bibinfo {year} {1996})}\BibitemShut {NoStop}%
\bibitem [{\citenamefont {Wu}\ \emph {et~al.}(2022{\natexlab{b}})\citenamefont
  {Wu}, \citenamefont {Zhou}, \citenamefont {Zhou}, \citenamefont {Wang},\ and\
  \citenamefont {Ji}}]{JW}%
  \BibitemOpen
  \bibfield  {author} {\bibinfo {author} {\bibfnamefont {L.}~\bibnamefont
  {Wu}}, \bibinfo {author} {\bibfnamefont {L.}~\bibnamefont {Zhou}}, \bibinfo
  {author} {\bibfnamefont {X.}~\bibnamefont {Zhou}}, \bibinfo {author}
  {\bibfnamefont {C.}~\bibnamefont {Wang}}, \ and\ \bibinfo {author}
  {\bibfnamefont {W.}~\bibnamefont {Ji}},\ }\href
  {https://link.aps.org/doi/10.1103/PhysRevB.106.L081401} {\bibfield  {journal}
  {\bibinfo  {journal} {Phys. Rev. B}\ }\textbf {\bibinfo {volume} {106}},\
  \bibinfo {pages} {L081401} (\bibinfo {year}
  {2022}{\natexlab{b}})}\BibitemShut {NoStop}%
\bibitem [{\citenamefont {Grimme}\ \emph {et~al.}(2010)\citenamefont {Grimme},
  \citenamefont {Antony}, \citenamefont {Ehrlich},\ and\ \citenamefont
  {Krieg}}]{DFT-D3}%
  \BibitemOpen
  \bibfield  {author} {\bibinfo {author} {\bibfnamefont {S.}~\bibnamefont
  {Grimme}}, \bibinfo {author} {\bibfnamefont {J.}~\bibnamefont {Antony}},
  \bibinfo {author} {\bibfnamefont {S.}~\bibnamefont {Ehrlich}}, \ and\
  \bibinfo {author} {\bibfnamefont {H.}~\bibnamefont {Krieg}},\ }\href
  {https://doi.org/10.1063/1.3382344} {\bibfield  {journal} {\bibinfo
  {journal} {J. Chem. Phys.}\ }\textbf {\bibinfo {volume} {132}},\ \bibinfo
  {pages} {154104} (\bibinfo {year} {2010})}\BibitemShut {NoStop}%
\bibitem [{\citenamefont {Lou}\ \emph {et~al.}(2021)\citenamefont {Lou},
  \citenamefont {Li}, \citenamefont {Ji}, \citenamefont {Yu}, \citenamefont
  {Feng}, \citenamefont {Gong},\ and\ \citenamefont {Xiang}}]{PASP}%
  \BibitemOpen
  \bibfield  {author} {\bibinfo {author} {\bibfnamefont {F.}~\bibnamefont
  {Lou}}, \bibinfo {author} {\bibfnamefont {X.}~\bibnamefont {Li}}, \bibinfo
  {author} {\bibfnamefont {J.}~\bibnamefont {Ji}}, \bibinfo {author}
  {\bibfnamefont {H.}~\bibnamefont {Yu}}, \bibinfo {author} {\bibfnamefont
  {J.}~\bibnamefont {Feng}}, \bibinfo {author} {\bibfnamefont {X.}~\bibnamefont
  {Gong}}, \ and\ \bibinfo {author} {\bibfnamefont {H.}~\bibnamefont {Xiang}},\
  }\href {https://doi.org/10.1063/5.0043703} {\bibfield  {journal} {\bibinfo
  {journal} {J. Chem. Phys.}\ }\textbf {\bibinfo {volume} {154}},\ \bibinfo
  {pages} {114103} (\bibinfo {year} {2021})}\BibitemShut {NoStop}%
\bibitem [{\citenamefont {Li}\ \emph {et~al.}(2020)\citenamefont {Li},
  \citenamefont {Lou}, \citenamefont {Gong},\ and\ \citenamefont
  {Xiang}}]{LXY}%
  \BibitemOpen
  \bibfield  {author} {\bibinfo {author} {\bibfnamefont {X.-Y.}\ \bibnamefont
  {Li}}, \bibinfo {author} {\bibfnamefont {F.}~\bibnamefont {Lou}}, \bibinfo
  {author} {\bibfnamefont {X.-G.}\ \bibnamefont {Gong}}, \ and\ \bibinfo
  {author} {\bibfnamefont {H.}~\bibnamefont {Xiang}},\ }\href
  {https://iopscience.iop.org/article/10.1088/1367-2630/ab85df/meta} {\bibfield
   {journal} {\bibinfo  {journal} {New J. Phys.}\ }\textbf {\bibinfo {volume}
  {22}},\ \bibinfo {pages} {053036} (\bibinfo {year} {2020})}\BibitemShut
  {NoStop}%
\bibitem [{\citenamefont {Xu}\ \emph {et~al.}(2024)\citenamefont {Xu},
  \citenamefont {Yu}, \citenamefont {Wang},\ and\ \citenamefont {Xiang}}]{XCS}%
  \BibitemOpen
  \bibfield  {author} {\bibinfo {author} {\bibfnamefont {C.}~\bibnamefont
  {Xu}}, \bibinfo {author} {\bibfnamefont {H.}~\bibnamefont {Yu}}, \bibinfo
  {author} {\bibfnamefont {J.}~\bibnamefont {Wang}}, \ and\ \bibinfo {author}
  {\bibfnamefont {H.}~\bibnamefont {Xiang}},\ }\href {https://www.annualreviews.org/doi/abs/10.1146/annurev-conmatphys-032922-102353} {\bibfield  {journal} {\bibinfo
  {journal} {Annu. Rev. Condens. Matter Phys.}\ }\textbf {\bibinfo {volume}
  {15}},\ \bibinfo {pages} {85} (\bibinfo {year} {2024})}\BibitemShut {NoStop}%
\bibitem [{\citenamefont {Li}\ \emph {et~al.}(2023{\natexlab{b}})\citenamefont
  {Li}, \citenamefont {Xu}, \citenamefont {Liu}, \citenamefont {Li},
  \citenamefont {Bellaiche},\ and\ \citenamefont {Xiang}}]{NovelNiI2}%
  \BibitemOpen
  \bibfield  {author} {\bibinfo {author} {\bibfnamefont {X.}~\bibnamefont
  {Li}}, \bibinfo {author} {\bibfnamefont {C.}~\bibnamefont {Xu}}, \bibinfo
  {author} {\bibfnamefont {B.}~\bibnamefont {Liu}}, \bibinfo {author}
  {\bibfnamefont {X.}~\bibnamefont {Li}}, \bibinfo {author} {\bibfnamefont
  {L.}~\bibnamefont {Bellaiche}}, \ and\ \bibinfo {author} {\bibfnamefont
  {H.}~\bibnamefont {Xiang}},\ }\href
  {https://link.aps.org/doi/10.1103/PhysRevLett.131.036701} {\bibfield
  {journal} {\bibinfo  {journal} {Phys. Rev. Lett.}\ }\textbf {\bibinfo
  {volume} {131}},\ \bibinfo {pages} {036701} (\bibinfo {year}
  {2023}{\natexlab{b}})}\BibitemShut {NoStop}%
\bibitem [{\citenamefont {Hukushima}\ and\ \citenamefont
  {Nemoto}(1996)}]{PTMC2}%
  \BibitemOpen
  \bibfield  {author} {\bibinfo {author} {\bibfnamefont {K.}~\bibnamefont
  {Hukushima}}\ and\ \bibinfo {author} {\bibfnamefont {K.}~\bibnamefont
  {Nemoto}},\ }\href {https://journals.jps.jp/doi/abs/10.1143/JPSJ.65.1604}
  {\bibfield  {journal} {\bibinfo  {journal} {J. Phys. Soc. Jpn}\ }\textbf
  {\bibinfo {volume} {65}},\ \bibinfo {pages} {1604} (\bibinfo {year}
  {1996})}\BibitemShut {NoStop}%
\bibitem [{\citenamefont {Stiefel}(1952)}]{CG}%
  \BibitemOpen
  \bibfield  {author} {\bibinfo {author} {\bibfnamefont {E.}~\bibnamefont
  {Stiefel}},\ }\href@noop {} {\bibfield  {journal} {\bibinfo  {journal} {J.
  Res. Nat. Bur. Standards}\ }\textbf {\bibinfo {volume} {49}},\ \bibinfo
  {pages} {409} (\bibinfo {year} {1952})}\BibitemShut {NoStop}%
\end{thebibliography}
\end{document}